\documentclass[preprint,3p,fleqn]{elsarticle}

\usepackage[T1]{fontenc}

\usepackage{amsmath}
\usepackage{amssymb}
\usepackage{mathtools}
\usepackage{bm}
\usepackage{booktabs}
\usepackage{siunitx}

\usepackage{refcount}

\usepackage{xcolor}
\usepackage{algorithm}
\usepackage{algpseudocode}
\usepackage{lineno}
\usepackage[colorlinks=true,linkcolor=black]{hyperref}
\hypersetup{
  pdfauthor={Pozzi, Ambrosi, Pezzuto}
}

\journal{arXiv}

\newcommand{\tens}[1]{\mathsf{#1}}

\newcommand{\R}{{\mathbb R}}

\newcommand{\bX}{{\boldsymbol X}}
\newcommand{\bx}{{\boldsymbol x}}
\newcommand{\bu}{{\boldsymbol u}}
\newcommand{\ba}{{\boldsymbol a}}
\newcommand{\bn}{{\boldsymbol n}}
\newcommand{\bv}{{\boldsymbol v}}
\newcommand{\bp}{{\boldsymbol p}}

\newcommand{\mR}{{\mathcal R}}
\newcommand{\mRL}{\mR_{\text{L}^2}}
\newcommand{\mRH}{\mR_{\text{H}^1}}
\newcommand{\mRTV}{\mR_{\text{TV}}}

\newcommand{\F}{{\tens F}}
\renewcommand{\P}{{\tens P}}
\newcommand{\A}{{\tens A}}
\newcommand{\I}{{\tens I}}

\newcommand{\dd}{\mathrm{d}}
\newcommand{\GammaD}{\Gamma_\text{D}}
\newcommand{\GammaN}{\Gamma_\text{N}}
\newcommand{\eps}{\varepsilon}

\newcommand{\rev}[1]{#1}

\renewcommand{\linelabel}[1]{}

\begin{document}

\begin{frontmatter}

\title{Reconstruction of the local contractility of the cardiac muscle from deficient apparent kinematics}

\author[PoliTO]{G.~Pozzi}
\ead{giulia.pozzi@polito.it}
\affiliation[PoliTO]{organization={Dipartimento di Scienze Matematiche ``G.L.~Lagrange'', Politecnico di Torino},
            country={Italy}}

\author[PoliTO]{D.~Ambrosi}
\ead{davide.ambrosi@polito.it}

\author[UTN,USI]{S.~Pezzuto\corref{cor1}}
\affiliation[UTN]{organization={Laboratory of Mathematics for Biology and Medicine, Department of Mathematics, Università di Trento},
            country={Italy}}
\affiliation[USI]{organization={Euler Institute, Università della Svizzera italiana},
            country={Switzerland}}
\cortext[cor1]{Corresponding author}
\ead{simone.pezzuto@unitn.it}

\begin{abstract}
Active solids are a large class of materials, including both living soft tissues and artificial matter, that share the ability to undergo strain even in absence of external loads.  While in engineered materials the actuation is typically designed {\it a priori}, in natural materials it is an unknown of the problem. In such a framework, the identification of inactive regions in active materials is of particular interest. An example of paramount relevance is cardiac mechanics and the assessment of regions of the cardiac muscle with impaired  contractility. The impossibility to measure the local active forces directly suggests us to develop a novel methodology exploiting kinematic data from clinical images by a variational approach to reconstruct the local contractility \rev{of} the cardiac muscle.
\rev{By finding the stationary points of a suitable cost functional we recover the contractility map of the muscle.}
Numerical experiments, including severe conditions with added noise to model uncertainties, and data knowledge limited to the boundary, demonstrate the effectiveness of our approach. Unlike other methods, we provide a spatially continuous recovery of the contractility map \rev{without compromising the computational efficiency.} 
\end{abstract}


\end{frontmatter}

\section{Introduction}
\label{sec::intro}
Soft active solids are materials able to strain and exhibit tensional fields even in absence of external loads. This class of materials encompasses a wide variety of samples, ranging from living soft tissues, such as muscles \cite{riccobelli2019activation}, to the artificial materials widely employed in soft robotics, i.e., active gels \cite{chung2021magnetically}. While in artificial active materials the actuation is designed to target the desired features of the sample, in natural materials the activation is not known \textit{a priori}.
For natural active solids it is therefore important to formulate mathematical methods to determine the intensity map of the activity, or even regions of compromised activation.
A paradigmatic example of such complexity is the cardiac muscle, where the stress generated by the activation of the cardiomyocites (cardiac cells) depends both on the location of the cells in the cardiac wall and on \rev{the} orientation of the fibers \cite{tortora2014anatomy}.

From the point of view of mathematical modelling, the active behavior of active solids can be accounted for by an \textit{active stress} term, which is added to the first variation of the strain energy density characterizing the material (the standard contribution of stress of inert matter). Alternatively, a less common approach---known as \textit{active strain}---mathematically represents the activation as a microstructural distortion, i.e.~a change in the reference (relaxed) configuration \cite{ambrosi2012active}. 
Regardless \rev{of} the choice of the approach, the active term must be constitutively stated, it depends on the specific problem at hand and can be a function of the deformation itself. The paradigmatic example in this respect is the Frank-Starling law of cardiac mechanics. 

The estimation of passive and active parameters embedded in the constitutive laws of soft materials is a problem that has long been addressed by the scientific community \cite{guccione1991passive, mojsejenko2015estimating, regazzoni2022machine, cicci2023uncertainty}. Above all, over the past few decades, such family of problems, classified as inverse problems from the mathematical perspective, are playing a crucial role in the field of personalized medicine where, for obvious reasons, the determination of material parameters cannot be addressed by the usual engineering technology. A striking example confirming the importance of a correct evaluation of the model parameters is cardiac modeling, where mathematical models have proven to be powerful diagnostic tools if combined with clinical data assimilation, for example integrating \textit{in-vivo} MRI \cite{sack2016personalised} or elastography data \cite{lee2007theoretical, zervantonakis2007novel}.

In this scenario, the most important contribution leveraging a mathematical model in the attempt of estimating the material properties of the cardiac tissue can be traced back to the work of \citet{guccione1991passive}, followed many years later by \citet{augenstein2005method} and \citet{wang2009modelling}. To this aim, different approaches were proposed ranging from standard techniques, as the minimization of the least-squares differences \cite{moulton1995inverse}, to more heuristic methods, such as genetic algorithms \cite{nair2007optimizing, mojsejenko2015estimating}.
In the context of cardiac modeling, the estimation of active parameters is even more challenging and it becomes crucial in the presence of pathologies that affect the active contraction of the tissue. An example is the paramount importance of identifying possible infarcted regions, i.e., portions of the cardiac tissue where dead cardiomyocytes have lost their contractile ability and do not contribute to the power stroke of the heart.
In this respect, a pioneering contribution is due to \citet{sermesant2006cardiac}, who proposed a framework to estimate local contractility of the ventricular myocardium using a cardiac mathematical model fed by clinical MRI images and leveraging data assimilation procedures to estimate local contractility from given displacements. Later, \citet{sun2009computationally} proposed a gradient-free optimization approach to facilitate the convergence towards the global minimum of the problem. 
In the same vein, several other approaches have been proposed, based on data assimilation principles, to reconstruct regional values of key biophysical parameters to feed mathematical models of the beating heart. However, all these attempts suffer from a major limitation: the estimation is restricted to a small number of contractility parameters\rev{~\cite{asner}, possibly} defined on prescribed regions \cite{delingette2011personalization, chabiniok2012estimation,finsberg}. This feature is restrictive especially when high precision in defining the scar boundaries is required. An improvement in this respect was made by \citet{kovacheva}, \rev{where the authors propose a method for estimating the \emph{local} (nodal) active tension in a 3-D left ventricle.}

\begin{figure}
    \centering
    \includegraphics[width=\linewidth]{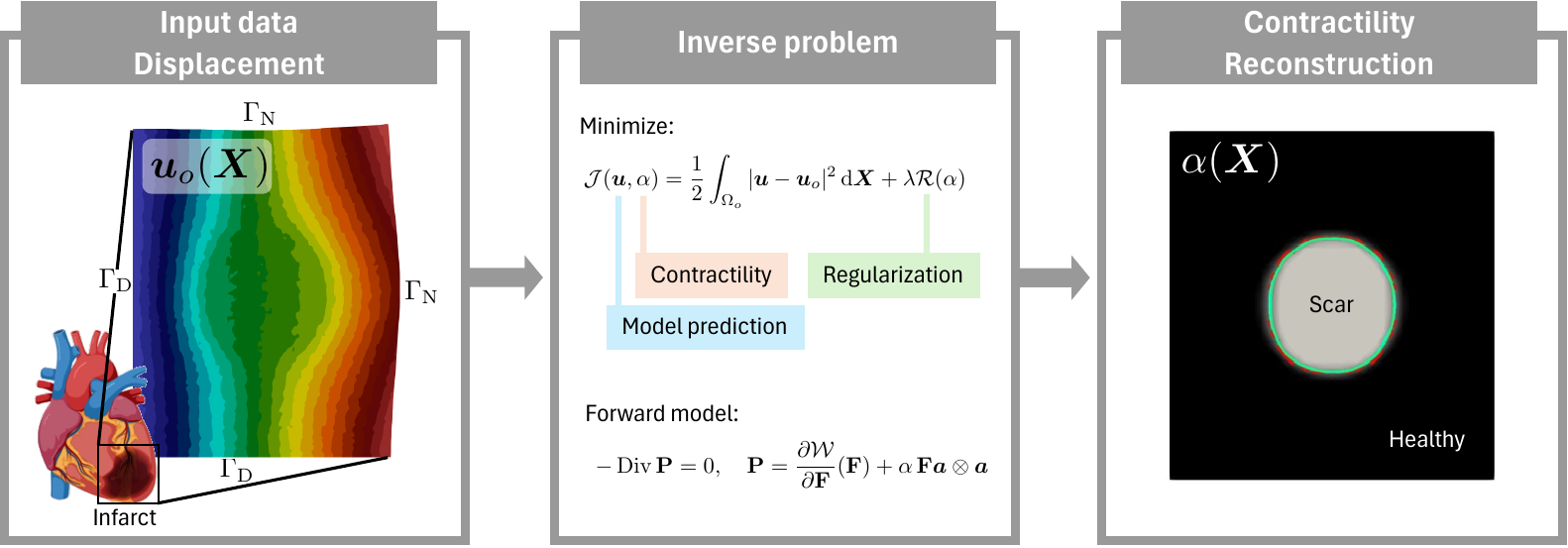} 
    \caption{Sketch of the process outline. We first generate artificial \textit{in-silico} data reproducing a 2D portion of the cardiac tissue presenting a scar---i.e.\ a region where the active contractility is impaired. For the given contractility field we solve the force balance equation and then perturb the solution (i.e. the displacement field) with a random noise to reproduce realistic data mimicking the output of imaging techniques. By an optimization technique, we solve the inverse problem to reconstruct the contractility function that minimizes the discrepancy between the simulated displacements and the artificial \textit{in-silico} data.}
    \label{fig:rational}
\end{figure}

\rev{The a}im of this paper is to propose a novel variational method to evaluate the activity of a contractile material and, in particular, to identify inert regions that possibly have a sharp boundary, on the basis of some recorded displacements. The application we have in mind is the determination of infarcted cardiac areas. Figure \ref{fig:rational} illustrates the main idea behind the proposed approach. In Section \ref{sec:mathmod} we address the inverse problem to determine an inhomogeneous material active stress on the basis of some knowledge of the given displacement field. On the basis of an \textit{a priori} knowledge of the constitutive law of the material and of the functional form of the active stress, we exploit a variational approach. After defining a suitable cost function, we look for the minimum of such functional and obtain the inverse problem, where the active stress is the unknown. The numerical solution of the inverse problem is discussed in Section~\ref{sec:numeth}. In Section \ref{sec:results},  we discuss the numerical results and perform a sensitivity analysis on varying the formulation or some parameters of the problem. 
Finally, in Section \ref{sec:conclusions} of the paper we discuss the obtained results comparing our approaches with the state-of-the-art methods.

\section{Mathematical statement of the problem}
\label{sec:mathmod}

\subsection{The forward problem}

Let us consider a continuous body defined in its reference configuration by the domain $\Omega \subset \R^d$, with $d=2,3$. We define the deformed configuration by a smooth map $$\Omega \ni \bX \mapsto \bx = \bx (\bX),$$
where $\bX$ and $\bx$ are the material and spatial coordinates, respectively. We also define the displacement field $\bu(\bX) = \bx(\bX) - \bX$ and the deformation gradient tensor $\F = \nabla \bx = \mathbf{I} + \nabla\bu$.

We assume that the body exhibits a hyperelastic behaviour and we constitutively provide an isotropic strain energy density per unit of reference volume $W(\F)$. As we are here mainly interested in methodological aspects, we keep the mechanical model at the minimum degree of complexity while retaining the essential features of inverse problems in finite elasticity. Accordingly, we assume the isotropic part to behave as a compressible neo-Hookean material~\cite{auricchio2013approximation}
\begin{equation}
	W(\F) = \frac{\mu}{2} \bigl(\F : \F -2 \log J - \rev{d} \bigr),
\label{eq::w_iso}
\end{equation}
where $J = \det\F$ and $\mu \ge 0$ is the shear modulus.
The strain energy in Eq.~\eqref{eq::w_iso} corresponds, in a small strain regime, to an isotropic linear material with shear modulus $\mu$ and zero Poisson's ratio. Volume variation are penalized logarithmically at large strains. All the calculations that follow can be easily extended to more complex strain energies. 

Since we assume the material to be active, the Piola stress tensor corresponding to Eq.~\eqref{eq::w_iso} must be complemented by the active stress component,
\begin{equation}
\rev{
\P_a = \alpha \, \F \ba \otimes \ba,
\label{eq:Pa}
}%
\end{equation}
where $\ba = \ba(\bX)$ is the unit vector parallel to the fiber orientation in the reference configuration. Again, the specific form of the active stress we introduce in \eqref{eq::piola} is driven by a sake of simplicity only: more complex functional forms could be successfully adopted in the same methodological framework. The  contractility $\alpha = \alpha(\bX)$ modulates the amplitude of the active stress.
\rev{The} total Piola reads as follows:
\begin{equation} 
	\P(\F,\alpha) = \frac{\partial W}{\partial \F} + \P_a  
	       = \mu (\F - \F^{-T})
           + \alpha \, \F \ba \otimes \ba.
\label{eq::piola}
\end{equation}

Overall, we impose the force balance
\begin{equation}
	      - \nabla \cdot \P = \mathbf{0},
\label{eq::fbalance}
\end{equation}
that, for the sake of simplicity, we equip with homogeneous boundary conditions
\begin{equation}
	      \P\bn \Big|_{\GammaN}  = \mathbf{0},
    \qquad  \bu \Big|_{\GammaD}  = \mathbf{0},
\label{eq::bcs}
\end{equation}
where $\GammaN$ and $\GammaD$, with $\rev{\bar{\Gamma}_\text{N}} \cup \rev{\bar\Gamma_\text{D}} = \partial \Omega $ \rev{and $\bar{\Gamma}_\text{N} \cap \GammaD = \GammaN \cap \bar{\Gamma}_\text{D} = \emptyset$}, denote the Neumann and the Dirichlet boundary, respectively.

In summary, the forward (or primal) problem consists in finding the displacement field $\bu(\bX)$ given an active stress $\alpha(\bX)$. We remark that, under suitable regularity assumptions on the domain and parameters, the existence of a solution to the problem~\eqref{eq::fbalance} is ensured by the polyconvexity of the strain energy~\eqref{eq::w_iso} and the active stress term~\cite{schroder2003invariant} \rev{supplemented by the growth conditions stated in \cite{ball1976convexity,hughes1983mathematical}}.

\subsection{The cost functional}
\label{subsec:cost}
Let \rev{us} now assume the contractility $\alpha(\bX)$ to be unknown and the displacement of the body $\bu(\bX)$ to be observable in a subregion $\Omega_o \subseteq \Omega$.

We thus aim at exploiting the knowledge of the observed displacements $\bu_o$ to reconstruct the contractility function. As so, we define the cost functional
\begin{equation}
\mathcal{J}(\bu,\alpha) = 
  \frac{1}{2} \int_{\Omega_{o}} \rev{|\bu - \bu_{o}|^2} \
  \:\dd\bX 
+ \lambda\mR(\alpha), 
\label{eq::functional}
\end{equation}
\rev{where $\mR(\alpha)$} is a regularization term, \rev{$\lambda$ is the regularization weight} and $(\bu,\alpha)$ \rev{is} such that Eq.~\eqref{eq::fbalance} and Eq.~\eqref{eq::bcs} are satisfied. Together with \rev{the} physical requirement that $\alpha(\bX) \ge 0$ for all $\bX\in\Omega$, we formulate the inverse problem as the following constrained optimization problem:
\begin{equation}
    \min_{\bu, \alpha} \mathcal{J}(\bu, \alpha), \quad\text{s.t.}\quad
    \left\{\begin{aligned}
    - \nabla \cdot \P &= 0, && \text{in $\Omega$} \\
    \P\bn  &= 0,   && \text{on $\GammaN$}, \\
      \bu  &= 0,   && \text{on $\GammaD$}, \\
    \alpha &\ge 0, &&  \text{in $\Omega$}. \\
    \end{aligned}\right.
\label{eq::minpb}
\end{equation}

The regularization term is necessary to cope with the ill-posedness of the inverse problem and recover a stable solution~\cite{lions1971}.
Typically, inverse problems lead to solutions that are unstable with respect to data perturbations, for instance due to noise in the measurements~\cite{engl1996regularization}.
In this work we analyze three common choices for regularization, namely
\begin{subequations}
\begin{align}
\label{eq::reg_L2} 
\mR_{\mathrm{L}^2}(\alpha) &= \frac{1}{2} \int_\Omega  \alpha^2 \,\dd\bX,    \\ 
\label{eq::reg_H1} 
\mR_{\mathrm{H}^1}(\alpha) &= \frac{1}{2} \int_\Omega |\nabla \alpha |^2 \,\dd\bX,\\   
\label{eq::reg_TV} 
\mR_\mathrm{TV}(\alpha) &= \int_\Omega \sqrt{\eps + |\nabla \alpha|^2} \,\dd\bX,  
\end{align}
\end{subequations}
where $\eps \ll 1$ in \eqref{eq::reg_TV} is a parameter, in our case set to \num{e-2}. The first two are Tikhonov-type regularization terms respectively of order zero for $\mR_{\mathrm{L}^2}$ and of order one for $\mR_{\mathrm{H}^1}$~\cite{budd2011regularization}. Equation~\eqref{eq::reg_TV} is an approximation of the total variation regularization~\cite{rudin1992nonlinear, chan2005recent}, well-known in the computer vision and imaging community. The choice of the regularization term is critical for imposing the desired regularity on $\alpha(\bX)$, as shown in Sec.~\ref{sec:results}.

\subsection{Necessary optimality conditions}

A minimizer of the constrained optimization problem~\eqref{eq::minpb} must satisfy a set of optimality conditions that we derive \rev{here by formally following the standard Lagrange method for optimal control problems~\cite{troltzsch2010optimal, manzoni2021optimal}}. First, we introduce a Lagrange multiplier $\bp(\bX)$ such that $\bp(\bX)=0$ for $\bX\in\GammaD$, we formally remove the equality constraints (the forward problem and the boundary conditions) through the following Lagrangian functional,
\begin{equation}
    \mathcal{L}(\bu,\alpha, \bp)
    = \mathcal{J}(\bu,\alpha) - \int_{\Omega} (-\nabla\cdot\P )\cdot \bp \: \dd\bX
    = \mathcal{J}(\bu,\alpha) + \int_{\Omega} \P : \nabla\bp \: \dd\bX,
    \label{eq::lagrangian}
\end{equation}
where we have used the Neumann boundary condition in the integration by parts. We recall that if $(\bu^*,\alpha^*,\bp^*)$ is a saddle point of the Lagrangian in Eq.~\eqref{eq::lagrangian}, then $(\bu^*,\alpha^*)$ is a minimizer of Eq.~\eqref{eq::minpb}. A necessary condition for $(\bu,\alpha,\bp)$ being a saddle point is that the first variation of the Lagrangian vanishes therein. Hence, we require that
\begin{equation}
\delta\mathcal{L}(\bu^*,\alpha^*,\bp^*) = 
  \frac{\partial\mathcal{L}}{\partial\bu}(\bu^*,\alpha^*,\bp^*)\,\delta\bu
+ \frac{\partial\mathcal{L}}{\partial\alpha}(\bu^*,\alpha^*,\bp^*)\,\delta\alpha
+ \frac{\partial\mathcal{L}}{\partial\bp}(\bu^*,\alpha^*,\bp^*)\,\delta\bp = 0,
\label{eq::varpb}
\end{equation}
for all admissible variations $(\delta\bu,\delta\alpha,\delta\bp)$.

To fix the ideas, in the following we consider $\mR_\mathrm{TV}$ regularization. The Lagrangian reads as follows:
\begin{equation}
    \mathcal{L}(\bu, \alpha, \bp) = 
    \frac{1}{2} \int_{\Omega_{o}} \rev{|\bu - \bu_{o}|^2} \: \dd{\bX}
    + \lambda \int_\Omega \sqrt{\eps + |\nabla\alpha|^2} \,\dd\bX 
    + \int_\Omega \P : \nabla\bp \: \dd\bX.
\end{equation}
The first variation follows from term-by-term differentiation of the above expression. We start with
\begin{equation}
\begin{split}
\frac{\partial\mathcal{L}}{\partial\bu}\delta\bu &= \int_{\rev{\Omega_o}} \rev{(\bu - \bu_o)}\cdot\delta\bu \, \dd\bX \\
&+ \int_\Omega \Bigl( \mu \bigl(\nabla \delta\bu - \F^{-T}(\nabla \delta\bu)^T\F^{-T}\bigr) + \alpha\, (\nabla \delta\bu)\ba \otimes \ba \Bigr): \nabla\bp \: \dd\bX = 0,
\end{split}
\label{eq::dJdu}
\end{equation}
for all admissible $\delta\bu$, where we used the differentiation rule for a generic tensor  $\A$: $\delta(\A^{-1}) = - \A^{-1}\delta\A \A^{-1}$. \rev{Here $\nabla\delta\bu= \delta \nabla \bu = \delta \F$.
After integration by parts of the second term at the right hand side, the arbitrariness of the independent increments $\delta \bu$ yields the following \emph{adjoint problem}:}
\begin{equation}
\left\{ \begin{aligned}
- \nabla\cdot \mathbb{C}(\rev{\F},\alpha)[\nabla\bp] &= \rev{(\mathcal{E}\bu_o - \bu)\chi_{\Omega_o}}, && \text{in $\Omega$,} \\
\mathbb{C}(\rev{\F},\alpha)[\nabla\bp]\bn &= \mathbf{0}, && \text{on $\GammaN$,}
\end{aligned} \right.
\label{eq::adjoint}
\end{equation}
where \rev{with $\chi_{\Omega_o}$ we denote the indicator function of $\Omega_o$ and $\mathcal{E}$ is the prologation operator from $\Omega_0$ to $\Omega$.}
Moreover, we have introduced the fourth-order Lagrangian tangent modulus
\begin{equation}
\mathbb{C}(\rev{\F},\alpha)[\A] = \mu \bigl(\A - \F^{-T}\A^T\F^{-T} \bigr) + \alpha\, \A \ba \otimes \ba.
\label{eq::Cten}
\end{equation}

Similarly, the variation in $\bp$,
\begin{equation}
\frac{\partial\mathcal{L}}{\partial\bp}\delta\bp 
= \int_\Omega \P : \nabla\delta \bp \: \dd\bX = 0,
\label{eq::dJdp}
\end{equation}
for all admissible $\delta\bp$, gives the \emph{primal problem}, corresponding to Eqs.~\eqref{eq::fbalance} and \eqref{eq::bcs}.

The last term corresponds to the variation with respect to the control,
\begin{equation}
\frac{\partial\mathcal{L}}{\partial\alpha}\delta\alpha 
= \lambda \int_\Omega \frac{\nabla \alpha \cdot \nabla\delta\alpha}{\sqrt{\eps + |\nabla\alpha |^2}} \,\dd\bX
+ \int_\Omega \bigl( \delta\alpha\,\F\ba\otimes\ba \bigr): \nabla\bp \: \dd\bX = 0,
\label{eq::dJda}
\end{equation}
for all admissible $\delta\alpha$.
After integration by parts, we obtain the \emph{optimality condition},
\begin{equation}
\left\{ \begin{aligned}
\lambda \nabla\cdot \Biggl( \frac{\nabla \alpha}{\sqrt{\eps + \|\nabla\alpha\|^2}} \Biggr) &= \nabla\bp:(\F\ba\otimes\ba), && \text{in $\Omega$,} \\
\frac{\nabla \alpha \cdot \bn}{\sqrt{\eps + \|\nabla\alpha\|^2}} &= 0, && \text{on $\partial\Omega$.}
\end{aligned} \right.
\label{eq::optc}
\end{equation}



The \emph{optimal control problem} collects Eqs.~\eqref{eq::fbalance}, \eqref{eq::bcs}, \eqref{eq::adjoint}, and \eqref{eq::optc} into a single non-linear system in the unknowns $\bu$, $\alpha$, and $\bp$.

\rev{
\paragraph{Remark 1}
\linelabel{ll:alpha}
The choice of a physically correct boundary conditions for $\alpha$ is a tricky issue. From a cardiovascular point of view, this an immaterial question: in a three dimensional cardiac muscle the active fibers are always tangent to the free boundary so that the active stress has no normal component. In the somehow artificial two--dimensional setting of the numerical simulations of this work, there is no ``correct'' answer, in our opinion. Our choice of Neumann boundary conditions is the weakest one; imposing Dirichlet boundary conditions would be definitely legitimate too but the choice of the boundary value is not obvious. In an active strain context this ambiguity does not apply~\cite{ambrosi2012active}.
}

\rev{
\paragraph{Remark 2}
When data are known on some portion of the boundary only, the integral at the right hand side of Eq.~\eqref{eq::functional} rewrites as a surface (line, respectively) integral to be calculated on some $\Gamma_o \subseteq \GammaN$:
\begin{equation}
\mathcal{J}(\bu,\alpha) = 
  \frac{1}{2} \int_{\Gamma_{o}} \rev{|\bu - \bu_{o}|^2} \,\dd S 
+ \lambda\mR(\alpha). 
\label{eq::functionalBC}
\end{equation}
The calculations in such a case are formally different from the ones illustrated above, but with the same rationale and we do not report them explicitly. As a main difference, the resulting adjoint problem has the same structure of Eq.~\eqref{eq::adjoint} but the discrepancy rewrites in the Neumann boundary conditions, as a boundary load:
\begin{equation}
\left\{ \begin{aligned}
- \nabla\cdot \mathbb{C}(\rev{\F},\alpha)[\nabla\bp] &= \mathbf{0}, && \text{in $\Omega$,} \\
\mathbb{C}(\rev{\F},\alpha)[\nabla\bp]\bn &= (\mathcal{E}\bu_o - \bu)\chi_{\Gamma_o}, && \text{on $\GammaN$.}
\end{aligned} \right.
\label{eq::adjointBC}
\end{equation}
}


\rev{
\paragraph{Remark 3}
\linelabel{ll:inequality}The inequality constraint $\alpha \ge 0$ in Eq.~\eqref{eq::minpb} was purposely avoided in the above derivation for the sake of simplicity. More precisely, Eq.~\eqref{eq::optc} should be written in the form of inequality as follows:
\begin{equation*}
\frac{\partial\mathcal{L}}{\partial\alpha}(\delta\alpha - \alpha^*) \ge 0,
\end{equation*}
for all admissible $\delta\alpha \ge 0$. This aspect is covered in depth in books for optimal control theory~\cite[\S 2.10]{troltzsch2010optimal}. From a numerical viewpoint, the constraint is handled with a projected gradient approach (see Sec.~\ref{sec:numeth}).
}

\subsection{The reduced functional}
\label{sec::Jred}

From a numerical viewpoint, it is convenient to introduce the following \emph{unconstrained} optimization problem:
\begin{equation}
    \min_{\alpha} \mathcal{J}_\text{red}(\alpha) 
    = \mathcal{J}( \bu(\alpha), \alpha ),
\label{eq::Jred}
\end{equation}
where $\bu(\alpha)$ is the solution operator associated with the forward problem. The reformulation of the minimization problem this way allows to decouple the primal, adjoint and optimality problems from a numerical point of view, giving the possibility to adopt different  numerical strategies for each of them. \rev{Effectively}, the functional \eqref{eq::Jred} \rev{has no equality constraints, thus it is suitable for a larger class} of methods for unconstrained optimization, such as quasi-Newton methods. Formally, first-order optimization methods require, for a given $\bar\alpha$, the value of the cost functional $\mathcal{J}_\text{red}(\bar\alpha)$ and its gradient. For the evaluation of $\mathcal{J}_\text{red}(\bar\alpha)$, it is sufficient to first solve the forward problem to find $\bar\bu = \bu(\bar\alpha)$, and then compute $\mathcal{J}(\bar\bu, \bar\alpha)$ from Eq.~\eqref{eq::functional}.

For the computation of the gradient, we observe that if $\bu$ and $\bp$ are respectively solution of the forward~\eqref{eq::fbalance}-\eqref{eq::bcs} and adjoint~\eqref{eq::adjoint} problems, then
\begin{equation}
\delta \mathcal{J}_\text{red}(\alpha)
= \frac{\partial\mathcal{J}_\text{red}}{\partial\alpha}\delta\alpha
= \lambda \int_\Omega \frac{\nabla \alpha \cdot \nabla\delta\alpha}{\sqrt{\eps + \|\nabla\alpha\|^2}} \,\dd\bX
+ \int_\Omega \bigl( \delta\alpha\,\F\ba\otimes\ba \bigr): \nabla\bp \: \dd\bX.
\end{equation}
It follows that the Riesz's representative (the gradient) of $\partial_\alpha J_\text{red}$, conveniently denoted by $\nabla J_\mathrm{red}(\bar\alpha)$, is the unique function such that
\begin{equation}
\int_\Omega (\nabla \mathcal{J}_\text{red}) \delta\alpha \:\dd\bX
= \lambda \int_\Omega \frac{\nabla \bar\alpha \cdot \nabla\delta\alpha}{\sqrt{\eps + \|\nabla\bar\alpha\|^2}} \,\dd\bX
+ \int_\Omega \bigl( \delta\alpha\,\bar\F\ba\otimes\ba \bigr): \nabla\bar{\bp} \: \dd\bX
\label{eq::grad}
\end{equation}
for all directions $\delta\alpha$, where $\bar\F = \F(\bar\bu)$ and $\bar{\bp}$ solves the adjoint problem. Equation \eqref{eq::grad} is important in the numerical implementation because $-\nabla\mathcal{J}_\text{red}$ provides a descent direction for the optimization algorithm.

\paragraph{Remark}
The presented framework is not limited to the specific choices for the strain energy density function, boundary conditions, and activation model. For a general strain energy $\mathcal{W}(\F)$, we replace the elasticity tensor in Eq.~\eqref{eq::Cten} with the following
\begin{equation*}
\mathbb{C}(\F,\alpha)[\A] = \frac{\partial^2\mathcal{W}}{\partial\F\partial\F}[\A] + \alpha\, \A \ba \otimes \ba.
\end{equation*}

For strictly incompressible materials, as it often happens on soft tissue biomechanics, the primal problem must be supplemented with the constraint $J = 1$ and the Piola stress tensor in Eq.~\eqref{eq::piola} modifies as follows:
\begin{equation}
	\P(\F,\alpha) = \frac{\partial\mathcal{W}}{\partial\F} - p J \F^{-T} + \alpha \, \F \ba \otimes \ba,
	\label{eq::P_p_inc}
\end{equation}
where $p$ is the hydrostatic pressure. The corresponding adjoint problem follows as above.


\section{Numerical methods}
\label{sec:numeth}

\subsection{Numerical discretization}

For the numerical discretization of the problem we employ the Finite Element Method (FEM). The state, adjoint, and control variables are approximated with piecewise linear elements on a triangular mesh $\mathcal{T}_h$ of the domain $\Omega_h$, where $h$ is the average edge size. We denote by $Q_h$ the space of linear functions and by $V_h$ the space of vector-valued linear functions satisfying homogeneous Dirichlet conditions on $\GammaD$:
\begin{align*}
    Q_h &= \bigl\{ q_h \colon \text{$q_h|_K \in \mathbb{P}_1$ for all $K \in \mathcal{T}_h$} \bigr\}, \\
    V_h &= \bigl\{ \bv_h \colon \text{$\bv_h|_K \in [\mathbb{P}_1]^d$ for all $K \in \mathcal{T}_h$ and $\bv_h|_{\GammaD} = \mathbf{0}$} \bigr\}.
\end{align*}
The discretized state and adjoint variables are denoted by $\bu_h\in V_h$ and $\bp_h\in V_h$, respectively, whereas the control is $\alpha_h \in Q_h$.
Therefore, the discretized problem is obtained from the variational problem \rev{of} Eq.~\eqref{eq::varpb}, that is: Find $(\bu_h,\alpha_h,\bp_h)\in V_h \times Q_h \times V_h$ such that
\begin{equation}
\delta\mathcal{L}(\bu_h,\alpha_h,\bp_h) =
  \frac{\partial\mathcal{L}}{\partial\bu_h}\delta\bu_h
+ \frac{\partial\mathcal{L}}{\partial\alpha_h}\delta\alpha_h
+ \frac{\partial\mathcal{L}}{\partial\bp_h}\delta\bp_h = 0,
\label{eq::varpb_num}
\end{equation}
for all $(\delta\bu_h, \delta\alpha_h, \delta\bp_h) \in V_h \times Q_h \times V_h$. The resulting variational problem corresponds to Eq.~\eqref{eq::dJdu}, \eqref{eq::dJda}, and \eqref{eq::dJdp}, precisely:
\begin{align}
\label{eq::state_h}
\text{(state problem)} 
\quad& \int_\Omega \P_h : \nabla \delta\bp_h = 0, \\
\label{eq::adjoint_h}
\text{(adjoint problem)}
\quad& \int_\Omega \mathbb{C}(\F_h,\alpha_h)[\nabla\bp_h] : \nabla \delta\bu_h + \int_{\Omega_o} \rev{(\bu_h - \bu_{o,h})} \cdot \delta \bu_h \,\dd\bX = 0, \\
\label{eq::opt_h}
\text{(optimality)}
\quad& \int_\Omega \lambda \frac{\nabla \alpha_h \cdot \nabla\delta\alpha_h}{\sqrt{\eps + \|\nabla\alpha_h\|^2}} \,\dd\bX
+ \int_\Omega \bigl( \delta\alpha_h\,\F_h\ba\otimes\ba \bigr): \nabla\bp_h \: \dd\bX = 0,
\end{align}
where $\F_h = \I + \nabla\bu_h$, $\P_h = \P(\F_h,\alpha_h)$, \rev{and $\bu_{o,h}$ is the interpolation (or the projection) onto $V_h$ of the data $\bu_o$}. Equations~\eqref{eq::state_h}-\eqref{eq::opt_h} form a sparse, non-linear system with $N_V^2 N_Q$ unknowns, where $N_V = \dim V_h$ and $N_Q = \dim Q_h$.

\subsection{Solution of optimization problem}

The numerical solution of Eqs.~\eqref{eq::state_h}-\eqref{eq::opt_h} is challenging. A standard Newton's method may not converge \cite{nocedal1999numerical}, unless the initial guess is already close enough to the true solution. Moreover, the resulting tangent system has a saddle-point block structure, which means that its solution requires dedicated preconditioners.
This approach notoriously leads to high computational costs.

Here, we opt for a solution strategy based on a quasi-Newton optimization method. In other words, instead of directly solving the non-linear system monolithically, we minimize the reduced functional \eqref{eq::Jred}:
\begin{equation}
\min_{\alpha_h \in Q_h} \mathcal{J}_\text{red}(\alpha_h) = \min_{\alpha_h \in Q_h} \mathcal{J}\bigl(\bu_h(\alpha_h), \alpha_h\bigr),
\label{eq::Jred_num}
\end{equation}
under the constraint that $\alpha_h \ge 0$ and where $\bu_h(\alpha_h)$ is the solution operator of the discretized state problem in Eq.~\eqref{eq::state_h}. Note that the optimization problem \eqref{eq::Jred_num} is finite dimensional, since $\dim Q_h = N_Q < \infty$. For the sake of simplicity, we use the notation $\alpha_h$ also for the vector of coefficients in $\mathbb{R}^{N_Q}$.

A quasi-Newton method also requires the gradient of the objective function, denoted by $\nabla\mathcal{J}_\text{red}(\alpha_h)$. Thanks to the derivation in Sec.~\ref{sec::Jred} and Eq.~\eqref{eq::opt_h}, we have the following problem for the gradient: Find $\nabla\mathcal{J}_\text{red}(\alpha_h) \in Q_h$ such that
\begin{equation}
\int_\Omega \nabla\mathcal{J}_\text{red} \delta\alpha_h \, \dd\bX
=
\int_\Omega \lambda \frac{\nabla \alpha_h \cdot \nabla\delta\alpha_h}{\sqrt{\eps + \|\nabla\alpha_h\|^2}} \,\dd\bX
+ \int_\Omega \bigl( \delta\alpha_h\,\F_h\ba\otimes\ba \bigr): \nabla\bp_h \: \dd\bX,
\label{eq::gradJ_h}
\end{equation}
for all $\delta\alpha_h \in Q_h$. Note that when the optimality condition \eqref{eq::opt_h} is met, we are at a stationary point of the reduced functional, since $\nabla\mathcal{J}_\text{red}(\alpha_h) = 0$. In summary, given a tentative control $\alpha_h$ the computation of the gradient requires: the current state $\bu_h = \bu_h(\alpha_h)$, obtained by solving the state problem \eqref{eq::state_h}; the adjoint variable $\bp_h(\bu_h,\alpha_h)$, obtained by solving the adjoint problem \eqref{eq::adjoint_h} given $\bu_h$ and $\alpha_h$; the solution of the projection problem \eqref{eq::gradJ_h}.
See Algorithm \ref{algo::J} and Algorithm \ref{algo::gradJ} for a summary.

\begin{algorithm}[tb]
\begin{algorithmic}[1]
	\Require $\alpha_h \geq 0$,
	\State compute $\bu_h$ by solving the \textit{state problem} \eqref{eq::state_h} with $\alpha_h$
    \State evaluate $\mathcal{J}(\bu_h, \alpha_h)$
\end{algorithmic}
\caption{Calculation of $\mathcal{J}_\text{red}(\alpha_h)$}
\label{algo::J}
\end{algorithm}

\begin{algorithm}[tb]
\begin{algorithmic}[1]
    \Require $\alpha_h \geq 0$,
    \State compute $\bu_h$ by solving the \textit{state problem} \eqref{eq::state_h} with $\alpha_h$
    \State compute $\bp_h$ by solving the \textit{adjoint problem} \eqref{eq::adjoint_h} with $(\bu_h(\alpha_h), \alpha_h)$
    \State compute $\nabla\mathcal{J}_{red}(\alpha_h)$ from \eqref{eq::gradJ_h}
    \end{algorithmic}
\caption{Calculation of $\nabla \mathcal{J}_{red}(\alpha_h)$}
\label{algo::gradJ}
\end{algorithm}


For the optimization problem we use the Limited-memory BFGS method with Bound constraints (L-BFGS-B). This method iteratively updates $\alpha_h$ starting from an initial guess $\alpha_h^{(0)}$ as follows:
\begin{equation}
\alpha_h^{(k)} = \alpha_h^{(k-1)} + \beta^{(k-1)} q^{(k-1)},
\label{eq::alpha_update}
\end{equation}
where $q^{(k-1)}$ is a \emph{descent direction} and $\beta^{(k-1)}$ is the \emph{step length} (or learning rate). The BFGS algorithm selects $q^{(k-1)}$ and $\beta^{(k-1)}$ such that
\begin{equation*}
\mathcal{J}_\text{red}(\alpha^{(k)}_h) < \mathcal{J}_\text{red}(\alpha^{(k-1)}_h),
\end{equation*}
for all $k=1,2,\ldots$ until a convergence criterium is met. The descent direction is computed as follows:
\begin{equation}
	B^{(k)} q^{(k)} = -\nabla \mathcal{J}_\text{red}(\alpha_h^{(k)}),
	\label{eq::LS-direction}
\end{equation}
where $B^{(k)}$ is a $N_Q \times N_Q$ symmetric, non-singular matrix.
Note that when $B^{(k)}$ is positive-definite, for sufficiently small $\beta$ we have that
\begin{equation*}
\begin{split}
\mathcal{J}_\text{red}(\alpha^{(k)}_h)
&= \mathcal{J}_\text{red}\bigl( \alpha^{(k-1)}_h + \beta^{(k-1)} q^{(k-1)} \bigr) \\
&= \mathcal{J}_\text{red}\bigl(\alpha^{(k-1)}_h - \beta^{(k-1)} (B^{(k-1)})^{-1} \nabla\mathcal{J}_\text{red}\bigr) \\
&\approx \mathcal{J}_\text{red}(\alpha^{(k-1)}_h) - \beta^{(k-1)} (B^{(k-1)})^{-1} \nabla\mathcal{J}_\text{red}\cdot \nabla\mathcal{J}_\text{red} <
\mathcal{J}_\text{red}(\alpha^{(k-1)}_h),
\end{split}
\end{equation*}
thus we ensure a decrease of the objective functional.
In a quasi-Newton method, the matrix $B^{(k)}$ is an approximation of the Hessian matrix of $\mathcal{J}_\text{red}(\alpha_h^{(k)})$, which ensures a super-linear convergence rate towards the local minimum. (When $B^{(k)}$ is the Hessian matrix at $\alpha_h^{(k)}$ we obtain the quadratically-convergent Newton's method, whereas for $B^{(k)} = I$ we have the gradient descent algorithm.) The BFGS method constructs such approximation by rank-one updates of an initial guess $B^{(0)}$ (usually the identity matrix), so that solving the linear system in Eq.~\eqref{eq::LS-direction} is simplified by the Sherman-Morrison formula. In the limited-memory variant of the BFGS method (L-BFGS), the approximation is truncated to a maximum number of rank-one updates, usually around 20.
The step length $\beta^{(k-1)}$ is obtained by a back\-tracking line search method, that is, for a given direction $q^{(k-1)}$, the algorithm finds the largest $\beta$ that ensures a sufficient reduction of the function
\begin{equation*}
f(\beta) \coloneqq \mathcal{J}_\text{red}\bigl( \alpha_h^{(k-1)} + \beta q^{(k-1)} \bigr).
\end{equation*}
Finally, the bound $\alpha_h \ge 0$ is handled by a projected gradient strategy, and the iterative process stops when the norm of the projected gradient is below a fixed tolerance. A detailed and complete description of the L-BFGS-B algorithm is given in \cite{byrd1995limited}.

\paragraph{Implementation aspects}
The problem has been implemented using the open-source software Firedrake~\cite{FiredrakeUserManual}, which uses the PETSc library as linear algebra back-end \cite{petsc-efficient, petsc-web-page}.
The variational formulation is computed by means of the automatic differentiation tool UFL~\cite{alnaes2014unified}.
We used the Firedrake \texttt{adjoint} module, \rev{based on the \texttt{pyadjoint} package \cite{mitusch2019dolfin},} for the solution of the optimization problem, which internally uses the L-BFGS-B method
as implemented in the \texttt{scipy} library~\cite{zhu1997algorithm}.
\rev{For the computation of the gradient, we do not consider the projection step in Alg.~\ref{algo::gradJ}, and instead use the co-vector corresponding to the right hand side of the problem. This is necessary because the \texttt{scipy} version of the L-BFGS-B method cannot be informed of the metric of the problem through a re-definition of the inner product. Since we are using uniform meshes, the impact of this choice on the performance of the method is limited.}
The code is available at the address: \url{https://github.com/pezzus/invscar}.

\section{Numerical experiments}
\label{sec:results}

\subsection{Parameters settings and \textit{in-silico} data generation}
\label{subsec::settings}

\begin{figure}[t]
	\centering
	\includegraphics[width=0.43\textwidth]{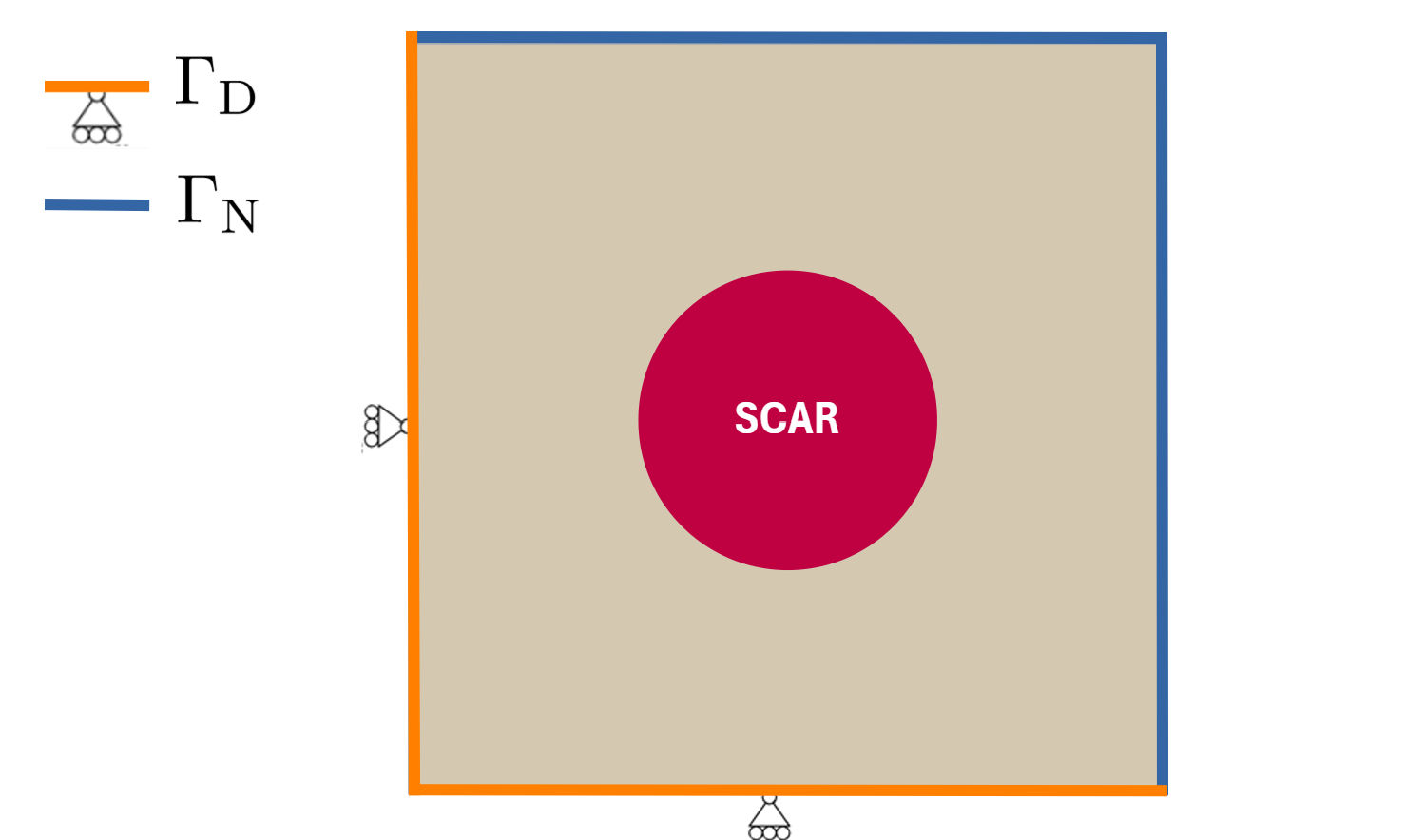}\hfill
    \includegraphics[width=0.27\textwidth]{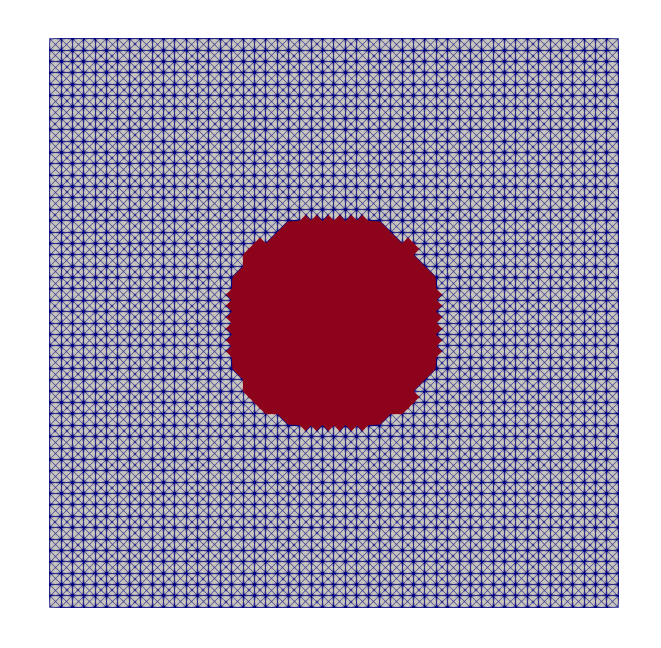}\hfill
   	\includegraphics[width=0.255\textwidth]{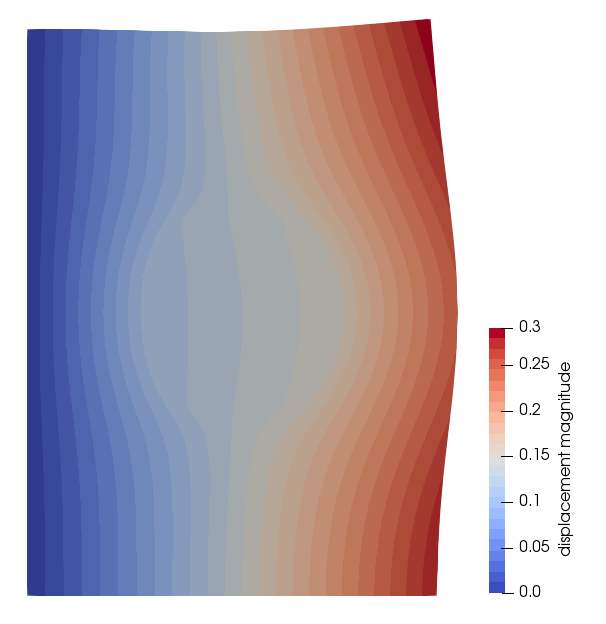}
	\caption{Sketch of the computational domain and of the boundary conditions (left); mesh of the reference configuration, the red spot representing the scar, i.e. the region where $\alpha = 0$ (center); \rev{ground truth: magnitude of the displacement represented in the deformed configuration (right)}.}
	\label{fig:mesh}
\end{figure}

Our reference numerical test is the reconstruction of a circular scar region centered in a planar, unit-square domain. For simplicity, all quantities are dimensionless. The geometrical setup is depicted in Fig.~\ref{fig:mesh}. We simulate the presence of a scar by imposing $\alpha = 0$ in a circular region of radius $r = 0.2$ centered in the box and we set $\alpha = 1$ elsewhere.
The elastic modulus $\mu = 1$ and the fibers are orientated horizontally. Boundary conditions apply as follows: only tangential  displacement is allowed at the bottom and left edge, free boundary elsewhere. 
The finite element discretization of the mesh is uniform and counts $10^4$ triangles.


To investigate the robustness of our approach, we generated \emph{in-silico} the reference solution $\bu_\text{ref}$ by adding a component-wise white noise $\boldsymbol{\eta}$ with zero mean and variance \num{e-6}, that is $\bu_o = \bu_\text{ref} + \boldsymbol{\eta}$. In such a way, the corresponding \textit{Signal to Noise Ratio} results
\begin{equation*}
    \text{SNR}_\text{dB} = 10 \log_{10}\frac{\Vert \bu_\text{ref} + \boldsymbol{\eta} \Vert^2}{\Vert \boldsymbol{\eta} \Vert^2} \simeq \SI{40}{\decibel},
\end{equation*}
on average, that corresponds to a relative perturbation of $1\%$ on the displacement field. \rev{Higher levels of noise are discussed in Sec.~\ref{sec:noise}.}

\rev{
The reconstruction of the contractility is obtained by minimization of the functional Eq.~\eqref{eq::functional} assuming observation on the full, that is, $\Omega_o = \Omega$, except for the boundary control  Sec.~\ref{sec:bcobs}, where we use Eq.~\eqref{eq::functionalBC} with $\Gamma_0 = \GammaN$. \linelabel{ll:initial_val}The initial guess is $\alpha=1$ everywhere. For the solution of the forward problem, we use the Newton's method with the direct solver MUMPS for the tangent problem~\cite{MUMPS:1} and we solve the adjoint problem with the same linear solver. The initial guess for the inner Newton's iteration is the previous solution of the optimization loop. Unless otherwise stated, we always consider the optimal regularization parameter for the solution of the inverse problem.}

\subsection{The effect of the regularization term}
\label{subsec::generalres}

Our first test is to solve the optimization problem~\eqref{eq::opt_h} for different regularization methods and at variance of the regularization parameter $\lambda$, all the other quantities being kept fixed.

\begin{figure}[t]
	\centering
	\includegraphics[width=0.33\linewidth,trim=25 15 10 10,clip]{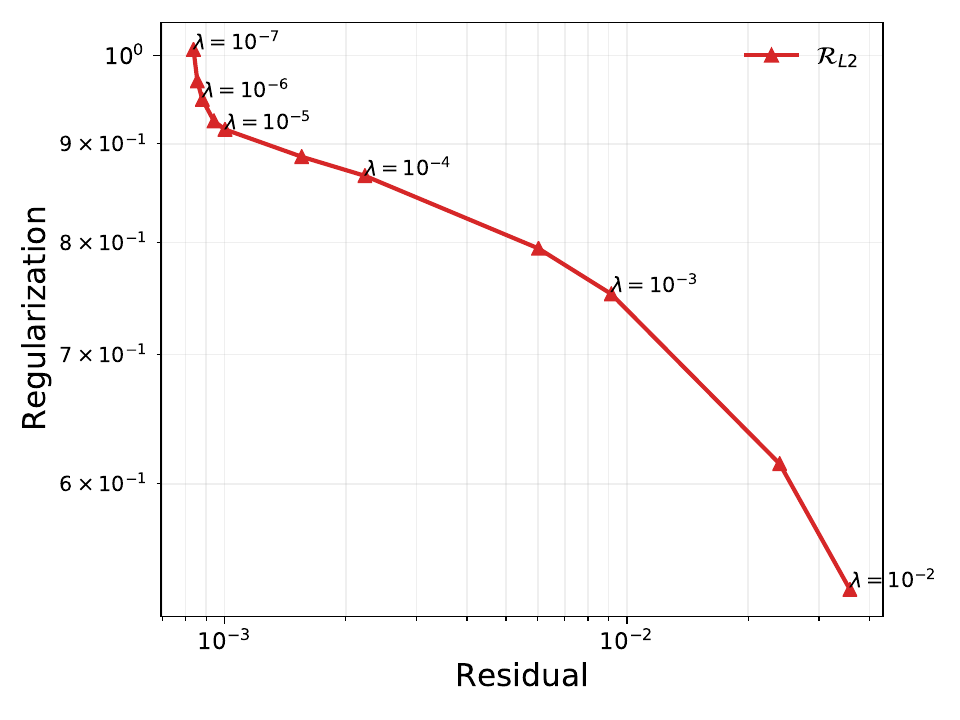}%
	\includegraphics[width=0.33\linewidth,trim=25 15 10 10,clip]{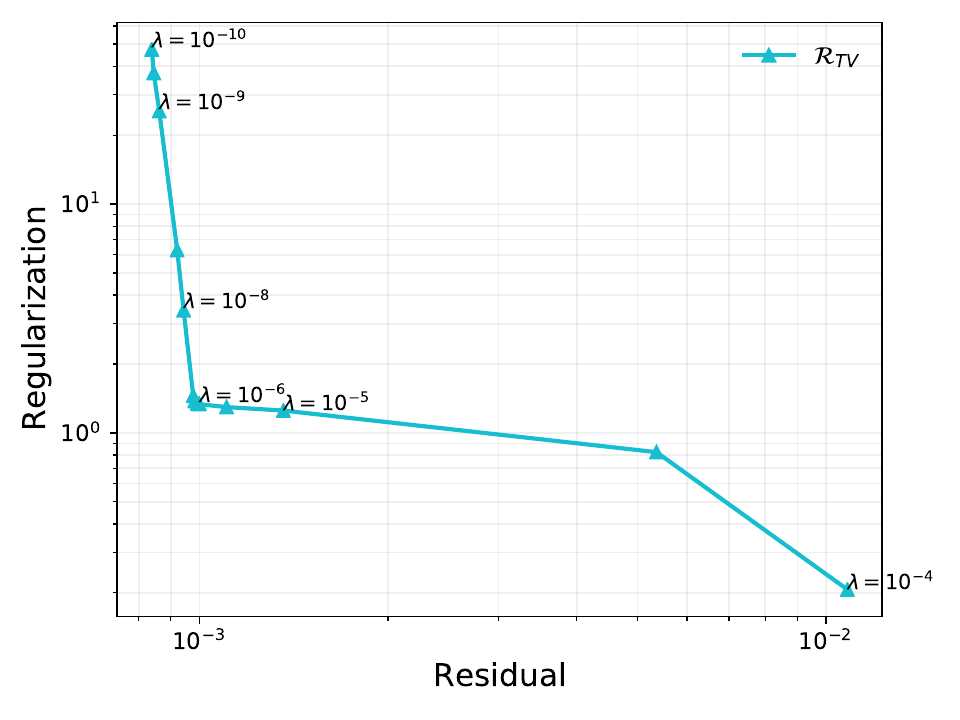}%
	\includegraphics[width=0.33\linewidth,trim=25 15 10 10,clip]{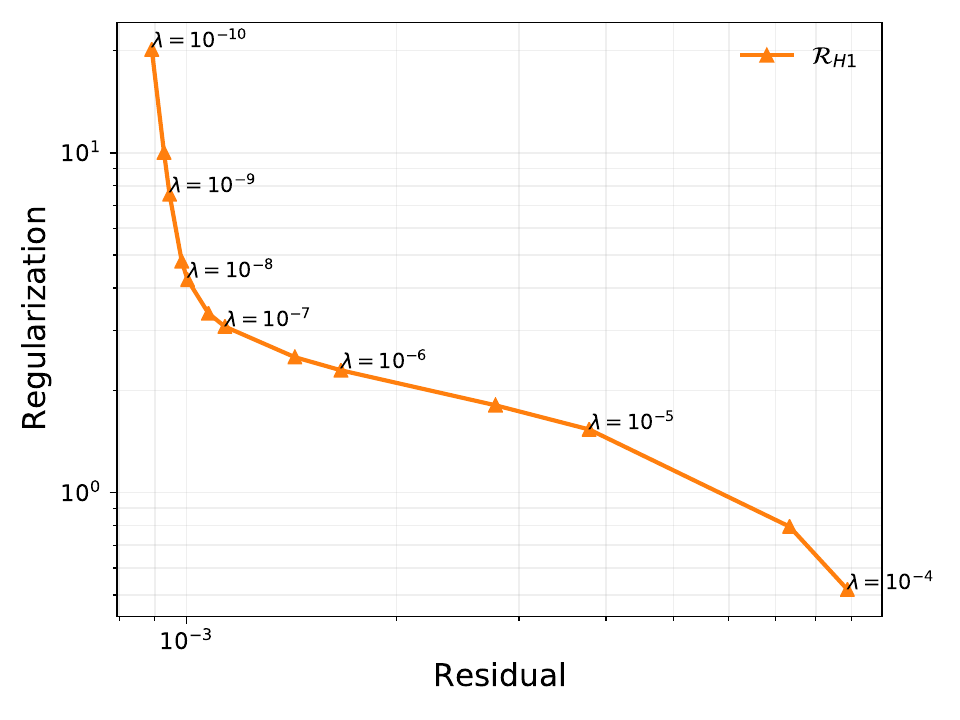}
	\caption{Pareto front for the optimal regularization weight. L-curves in logarithmic scale for three distinct regularizing terms in the cost functional: $\mathcal{R}_{L2}$ (left), $\mathcal{R}_{TV}$ (center) and $\mathcal{R}_{H1}$ (right). Triangular markers denote the simulated data for a given value of the parameter $\lambda$.}
	\label{fig::Lcurve}
\end{figure}

To identify the optimal value for $\lambda$, we computed the Pareto curve, \rev{which plots} the discrepancy versus the magnitude of the regularization, for each approach (see plot in Fig.~\ref{fig::Lcurve}). \linelabel{ll:pareto_cost}\rev{Each point on the curve corresponds to a solution of the inverse problem. Here, we select 11 values of $\lambda$ in logarithmic scale for $\mRH$ and $\mRTV$ regularization, and 13 values for $\mRL$.} The obtained curves are L-shaped where, according to the Pareto optimality principle, the optimal values correspond to the ``elbow'': that is $\lambda_\text{L2}^\text{opt} \approx \num{5e-5}$,
$\lambda_\text{TV}^\text{opt} \approx \num{e-6}$, and
$\lambda_\text{HI}^\text{opt} \approx \rev{\num{5e-8}}$.

\begin{figure}[t] 
	\centering
	\begin{tabular}{c c c c l}
	& & & \\
		\raisebox{5em}{$\mRL$} & 
		\includegraphics[width=0.22\linewidth]{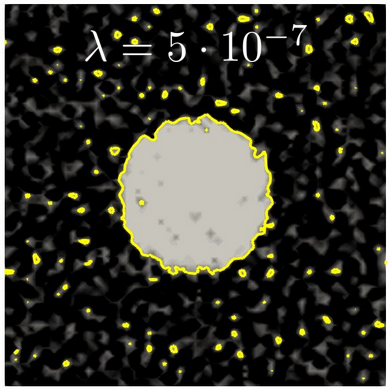} & 
		\includegraphics[width=0.22\linewidth]{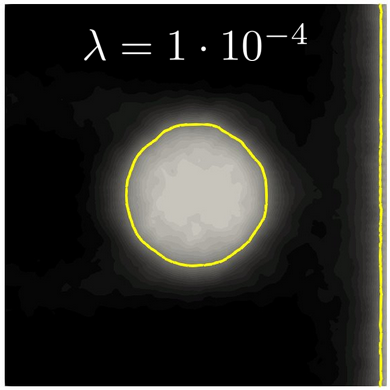} &
		\includegraphics[width=0.22\linewidth]{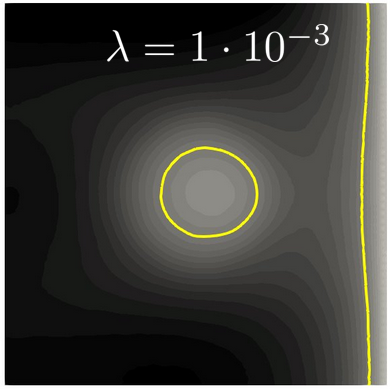} & \\
%
		\raisebox{5em}{$\mRTV$}  & 
		\includegraphics[width=0.22\linewidth]{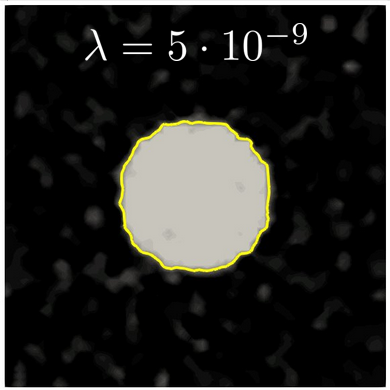} &  
		\includegraphics[width=0.22\linewidth]{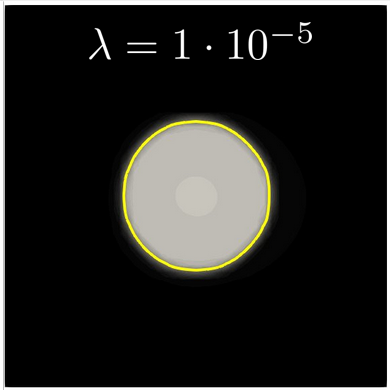} &
		\includegraphics[width=0.22\linewidth]{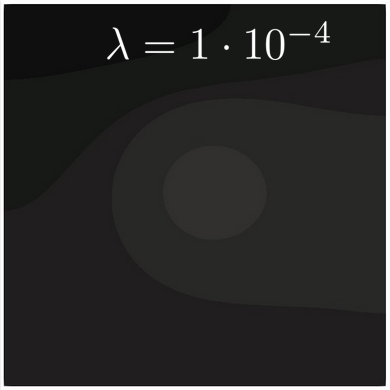} & \\
%
		\raisebox{5em}{$\mRH$} & 
		\includegraphics[width=0.22\linewidth]{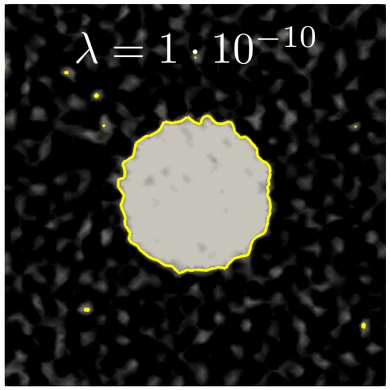} &  
		\includegraphics[width=0.22\linewidth]{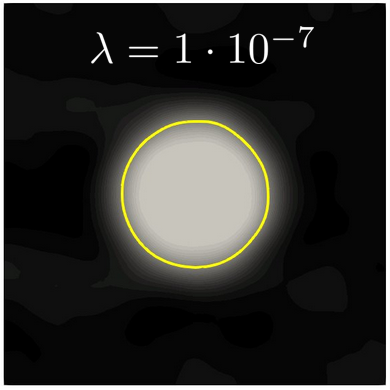} &
		\includegraphics[width=0.22\linewidth]{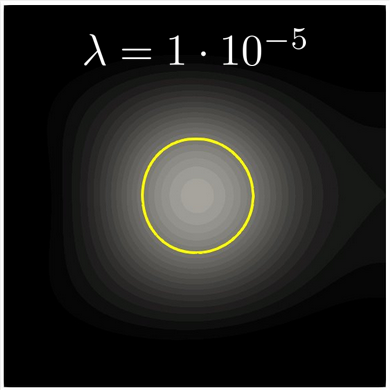} &
		\includegraphics[width=0.025\linewidth]{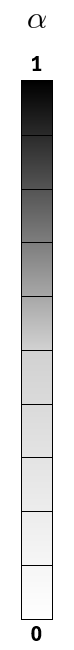}
	\end{tabular}
	\caption{
	Contractility reconstruction. Scar profiles reconstructed with the L-BFGS method for three distinct regularizing terms in the cost functional: $\mathcal{R}_\text{L2}$ (top row), $\mathcal{R}_\text{TV}$ (central row) and $\mathcal{R}_\text{H1}$ (bottom row). In each row, the relative weight $\lambda$ of the regularization is increased from left to right. The thin yellow contours represent the isolines $\alpha = 0.4$. To better highlight the discontinuous profile of the contractility function, the value of $\alpha$ was rescaled to the range $[0,1]$.}
	\label{tab::regularization}
\end{figure}

All the regularizing strategies exhibit a similar pattern, as shown in Fig.~\ref{tab::regularization} and 
the correct circular scar can be identified for some optimal value of $\lambda$.
For small values of $\lambda$, we notice the presence of spurious oscillations around the scarred area, more pronounced in L2 than H1 and TV regularizations. The noisy reconstruction for small $\lambda$ is a symptom of the ill-posedness of the inverse solution, because a small perturbation in the data $\bu_o$ corresponds to a large perturbation in the reconstruction.

For large values of $\lambda$, the contractility function $\alpha$ is smoother, blurring the boundary of the scar. This is particularly apparent with TV regularization, where for large $\lambda$ the scar cannot be detected.

With the optimal $\lambda$, the TV reconstruction matches very closely the ground truth, whereas the H1 and L2 reconstructions have smoother boundaries. The L2 reconstruction is also more prone to spurious oscillations. We also observe that the L2 reconstruction has a boundary layer artifact, located on the vertical free boundary (the right one) and particularly pronounced for large $\lambda$. We further investigated the presence of this artifact in \ref{appendix::A}, where we show that its position depends on the fiber direction and the boundary conditions.

The sharp approximation of the TV approach could be expected, because the TV regularization promotes piecewise-constant solutions and our choice of ground-truth solution falls in this class. In real case scenarios it might happen that the scar is surrounded by a smooth border characterized by remodelled tissue with a partial loss of contractility, thus the TV approach might incorrectly favor non-physiological reconstruction.

\begin{figure}[t]
	\centering
	\includegraphics[width=0.42\textwidth]{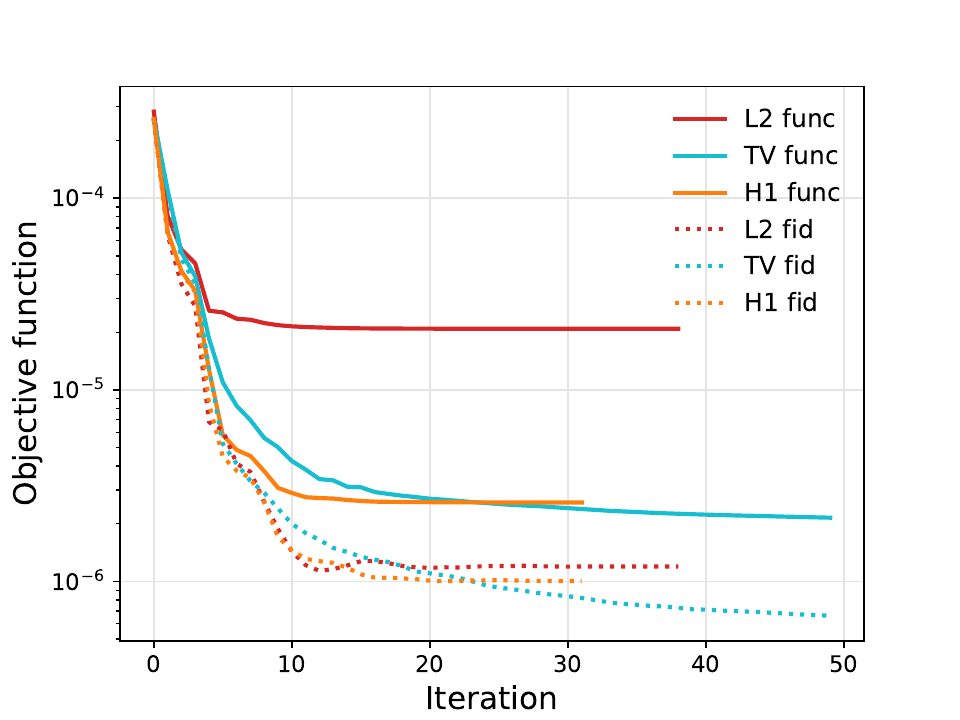}\hspace{1em}
    \includegraphics[width=0.42\textwidth]{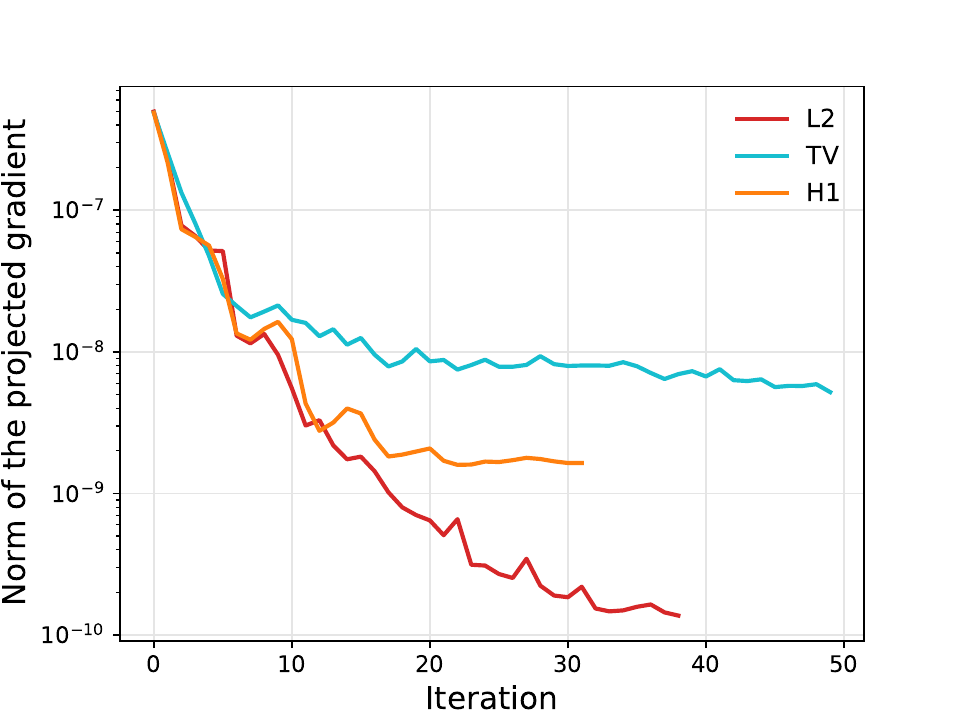}
	\caption{Convergence history for distinct regularizing terms. (Left) decay of the objective function (continuous lines), i.e. the cost functional, and of the fidelity term (dotted lines) over the iterations; 
	(right) norm of the projected gradient over the iterations.
    In both figures, the curves where obtained by setting $\lambda = \lambda_\text{opt}$ for each of the tested regularizing terms: $\mRL$ (red), $\mRTV$ (blue) and $\mRH$ (orange). \rev{The curve for TV is truncated to 50 iterations, although it requires more than 300 iterations to reach the desired tolerance.}}
	\label{fig:cvg_history}
\end{figure}

In Figure \ref{fig:cvg_history} we plot the convergence history obtained by different regularizing methods. In all cases, a good accuracy is achieved in less than 50 BFGS iterations. In fact, the convergence is very fast for L2 and H1 regularizations. For the TV approach the algorithm requires approximately 300 iterations to converge, although 20-30 iterations are already sufficient for \rev{a} good reconstruction. \linelabel{ll:tv_cost}The slow convergence rate of the TV approach is known in the literature, and it is due to the lack of regularity in the limit $\varepsilon\to 0$ in Eq.~\eqref{eq::reg_TV}. \rev{Here, the norm of the projected gradient, which is used as a convergence criterion in the L-BFGS-B method, decreases very slowly after the initial iterations.}

\linelabel{ll:cost}
\rev{The number of iterations needed to achieve convergence is quite independent with respect to the mesh size. Testing the $\mRH$ case with a mesh resolution from $25\times 25$ to $200\times 200$ (quadrilateral elements), the number of iterations varied from 27 to 34. The overall computational cost (on an Apple MacBook Pro M1 with 32 GB of memory, serial run) increased from \SI{0.67}{\sec} to \SI{37}{\sec}, due to the larger size of the forward and adjoint problems.
}

\rev{In summary, the H1 approach is the most efficient in terms of computational cost, convergence rate, and it can overcome the accuracy issues related to the presence of oscillations and artifacts at the boundary of the domain, where the L2 approach fails. On contrary, the TV approach is very accurate yet costly. 
}

\subsection{Sensitivity analysis: uncertainty and bias}
\label{subsec::SA}
In this section we perform a sensitivity analysis of the method assuming different amount of noise on the measured data and uncertainty on the model parameters. An effective inverse method should ideally weakly depend on parameters that cannot be directly measured. As a consequence, 
the ground truth and solution of the state equation in the inverse problem are different. \linelabel{ll:uq}\rev{However, we remark that our aim is not much to quantify the uncertainty in the computed fields,   ~\cite{Gander2021UQ, Quaglino2019MFMC}, but rather to assess the numerical stability of the algorithm due to specific modeling assumptions and to the presence of correlated noise in the input data.} 
Unless specified, all the reported results are obtained for the optimal value of the parameter $\lambda$.

\paragraph{Sensitivity to fiber direction}

\begin{figure}
	\centering
		
	\includegraphics[width=0.33\textwidth,trim=10 0 45 40,clip]{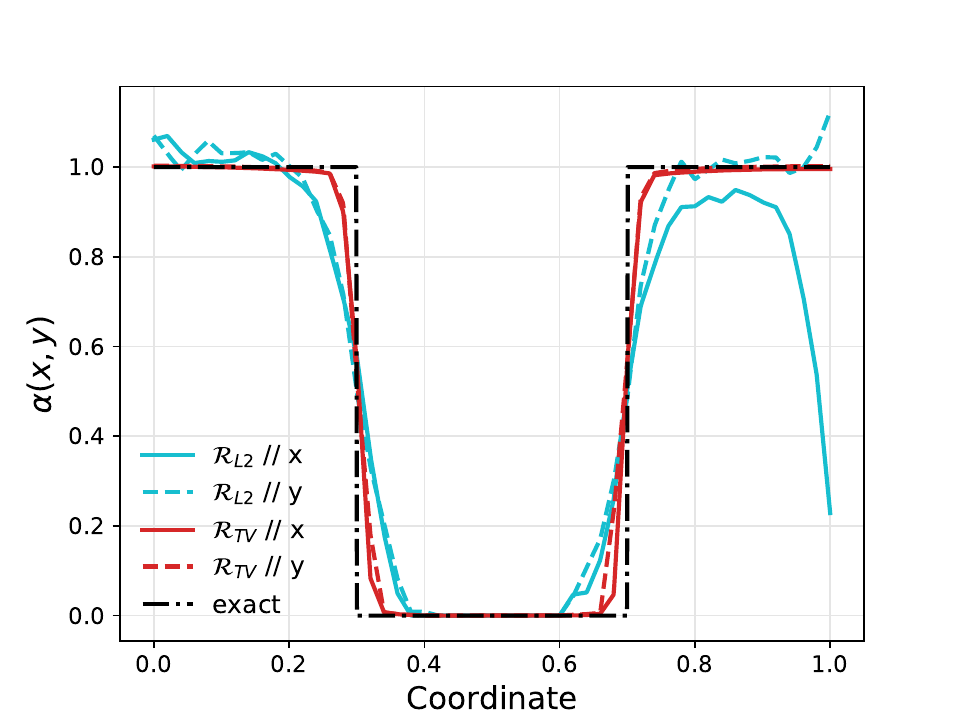}%
	\includegraphics[width=0.33\textwidth,trim=10 0 45 40,clip]{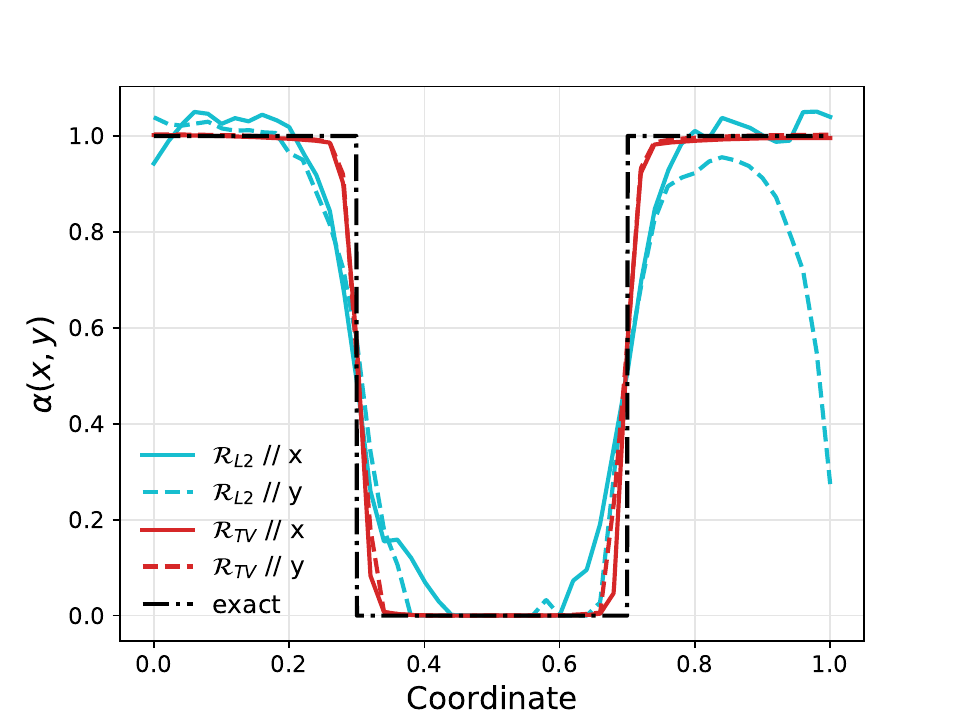}%
	\includegraphics[width=0.33\textwidth,trim=10 0 45 40,clip]{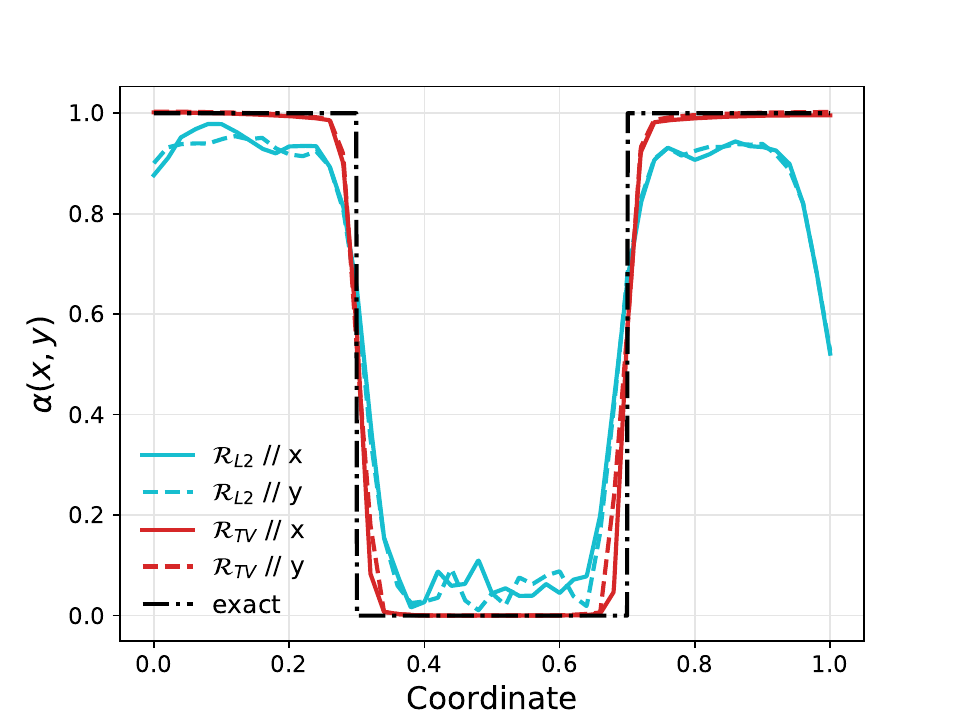} \\
 
    \vspace{1em}
	\hspace{6.5mm}\includegraphics[width=0.28\textwidth]{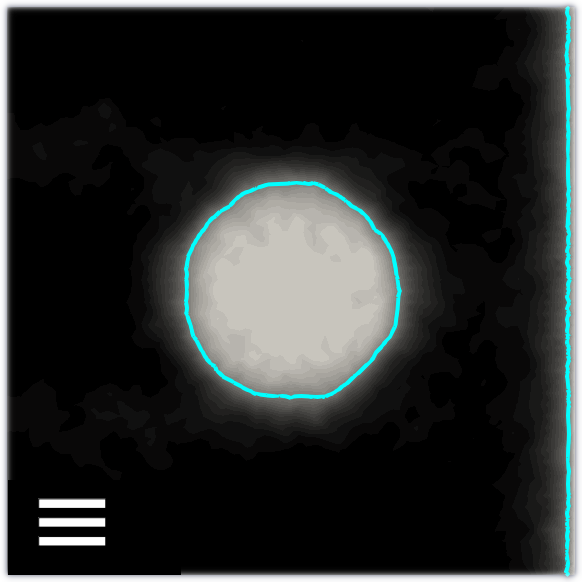}%
	\hspace{7mm}\includegraphics[width=0.28\textwidth]{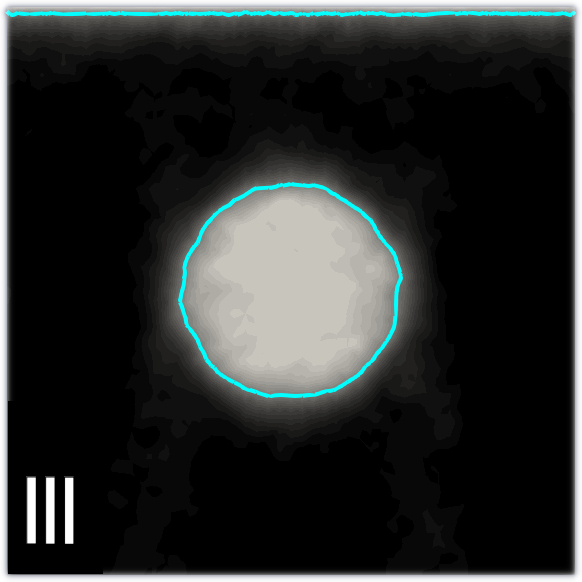}%
	\hspace{7.5mm}\includegraphics[width=0.28\textwidth]{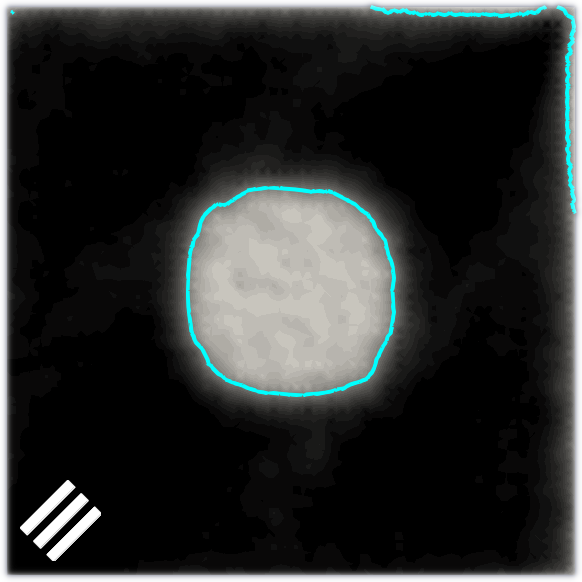}%
    \includegraphics[width=0.03\textwidth]{img/legenda_alpha_vert.PNG}\\
		
	\caption{Sensitivity of the model to the spatial fibers orientation. Three distinct fibers orientation are tested: fibers parallel to $x$-axis (left) parallel to $y$-axis (center) and fibers whose direction forms an angle of $\ang{45}$ with the $x$-axis (right).
    (Top row) Comparison among the exact profile of $\alpha(x,y)$ and its numerical reconstructions, along the $x$- and $y$-direction, obtained with both with $\mRL$ and $\mRTV$; (bottom row) scar profiles obtained from the numerical simulations (with $\mRL$) where cyan contours depicting isolines $\alpha = 0.4$. Fibers orientation is depicted in the bottom left. The values of $\alpha$ were rescaled to the range $[0,1]$ to better highlight the discontinuous profile of the contractility function. For each choice of the regularizing term, we set $\lambda = \lambda_\text{opt}.$}
	\label{fig::fib_orient}
\end{figure}

We first test the sensitivity of the inverse method to fibers orientation by L2 regularization by three distinct arrangements of the fibers while keeping the same boundary conditions. 
In Fig.~\ref{fig::fib_orient} we show that the $\mRTV$ and $\mRH$ regularization can accurately reconstruct the scar position, irrespective of the fiber direction. The $\mRL$ regularization suffers from a significant inaccuracy in the reconstruction of the contractility $\alpha(x,y)$ close to the stress-free boundary orthogonal to fibers. This issue is not present for the other regularization, because they can penalize sharp variations of $\nabla \alpha(x, y)$. We suspect that the anomalous behavior associated to $\mRL$ roots is related to the observability of the system (see \ref{appendix::A}).

\paragraph{Sensitivity to material parameters}
As a second test we include noise in the ground truth perturbing either the elastic modulus $\mu$ or the fiber orientation $\theta$ (where $\theta$ is s.t. $\ba =\left[\cos \theta, \sin\theta\right]$).
The ultimate goal of this analysis is to test the robustness of the proposed approach against observations generated by non-smooth and spatially-varying parameters. In fact, this will produce a spatially-correlated (colored noise) in the data.

\begin{figure}
	\centering
    \includegraphics[width=0.26\textwidth]{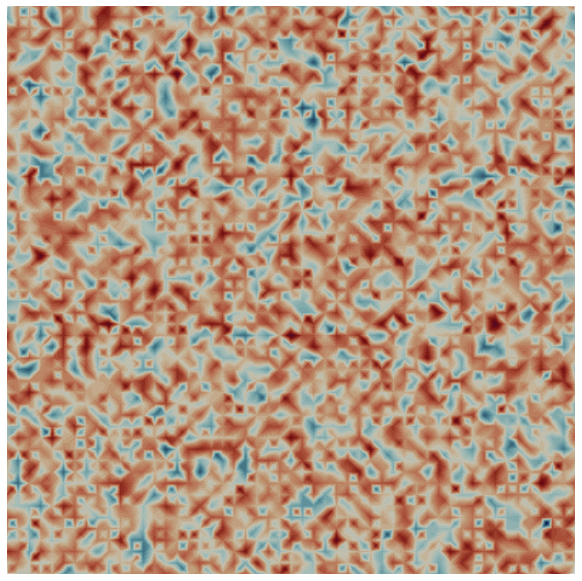}%
    \includegraphics[width=0.0335\linewidth]{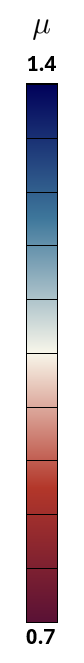}%
    \hfill
    \includegraphics[width=0.26\textwidth]{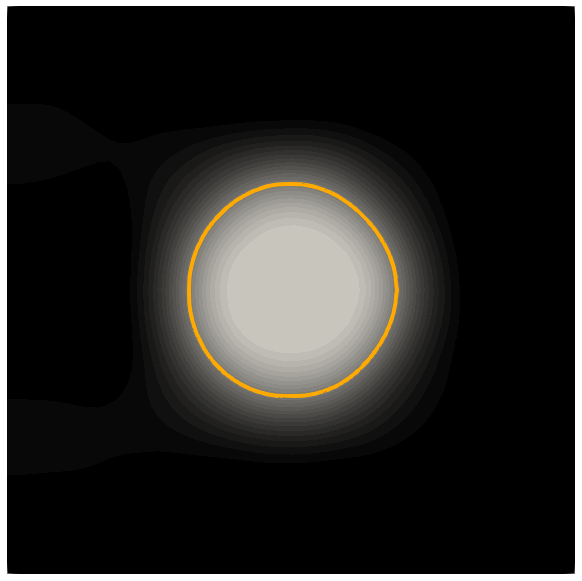}%
    \includegraphics[width=0.03\linewidth]{img/legenda_alpha_vert.PNG}%
    \hfill
    \includegraphics[width=0.34\textwidth,trim=30 0 40 40,clip]{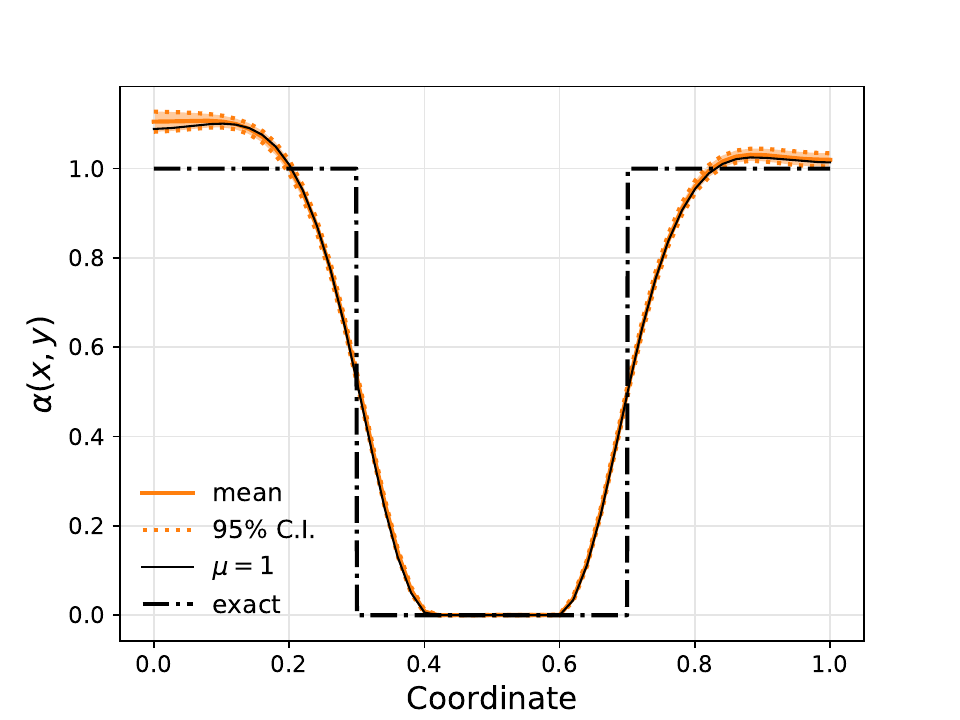}
    \\[1em]
    \includegraphics[width=0.26\textwidth]{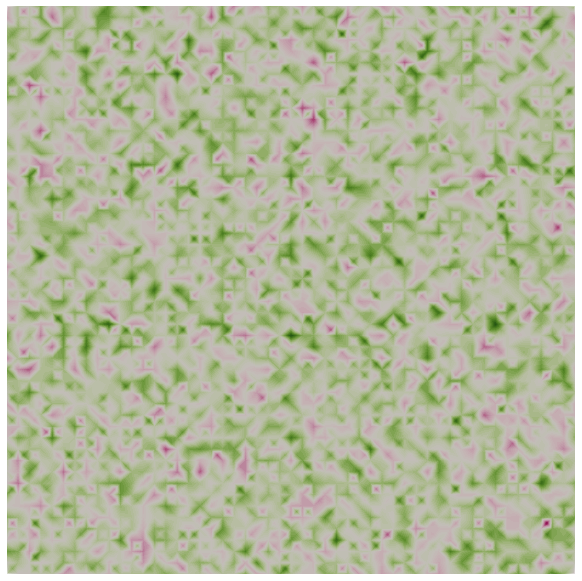}%
    \includegraphics[width=0.029\linewidth]{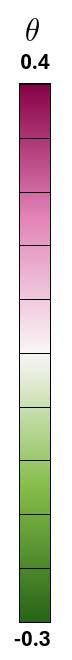}%
    \hfill
    \includegraphics[width=0.26\textwidth]{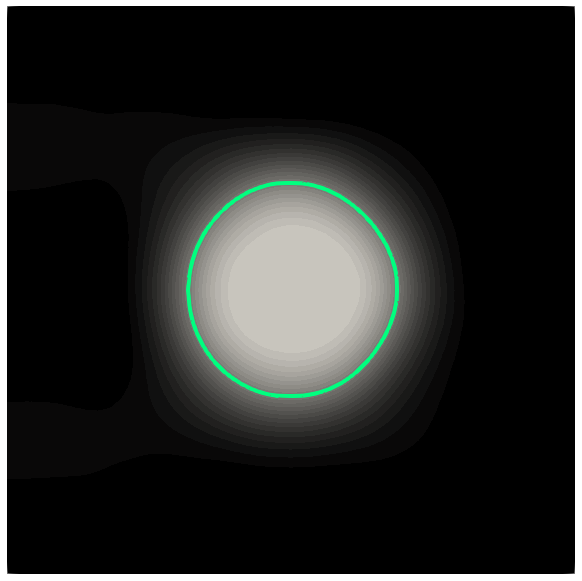}%
    \includegraphics[width=0.03\linewidth]{img/legenda_alpha_vert.PNG}%
    \hfill
    \includegraphics[width=0.34\textwidth,trim=30 0 40 40,clip]{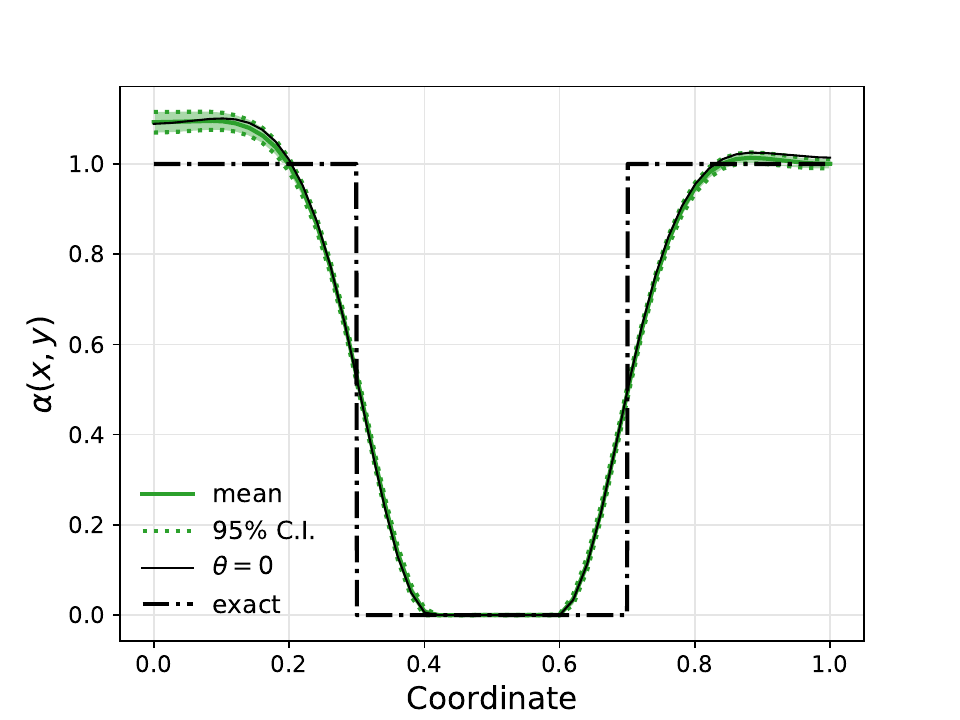}

    \caption{\rev{Robustness of the algorithm to perturbations of the parameter $\mu$ (top row) and fiber angle $\theta$, in radians (bottom row). (Left) A realization of the perturbed coefficient. (Middle) The reconstructed scar profile with $\mRH$ and $\lambda = \lambda^\text{opt}_\text{H1}$). Isolines $\alpha = 0.4$ are depicted by contours.
    (Right) Estimated contractility for constant parameter (black line) superimposed on the $95\%$-confidence interval (shaded area) for the mean contractility along the $x$-direction obtained from the Monte Carlo simulation ($N = 30$).}}
	\label{fig::mu_perturb}
\end{figure}

We first construct a $95\%$-confidence interval for the reconstructed contractility by using a Monte Carlo approach and generating the observed data $\bu_o$ from $\rev{N=30}$ different random perturbations of the elastic modulus $\mu$. Specifically, for each Monte Carlo step, we sample a different elastic modulus $\mu(\bX)$ from a Gaussian distribution centered in $\mu = 1$ and variance $\sigma^2=0.1$. The results of this test are reported in Fig.~\ref{fig::mu_perturb} \rev{(top row)}.

Secondly, we again adopt a Monte Carlo approach for generating a $95\%$-confidence interval for the reconstructed contractility where, in this second case, the observed data $\bu_o$ are generated from $N$ different random perturbations of local fiber orientation. This test aims at simulating the physiological disarray condition of the fibers in the heart \cite{finocchiaro2021arrhythmogenic}. As so, we generate different patterns of fibers by sampling the local value $\theta(\bX)$ from a \rev{n}ormal distribution centered in $\theta = 0$ and variance $\sigma^2 = 0.1$. The results of this test are reported in Fig.~\ref{fig::mu_perturb} \rev{(bottom row)}.

    

Overall, in both experiments the average reconstruction is close to the ground truth, although the contractility function is much smoother around the boundary of the scar. In terms of reconstruction uncertainty, we obtain narrow confidence intervals, thus supporting the robustness of the proposed approach in reconstructing the contractility even when the observed data are generated by non-smooth parameters.

\rev{
\paragraph{Sensitivity to an incorrect forward problem}

\linelabel{ll:incorrect_forward}
Next, we investigate the potential bias in the reconstruction of the scar when the forward model is biased or incorrect. First, we notice that we could rewrite the Piola tensor \eqref{eq::piola} as follows:
\begin{equation*}
\P = \mu \Bigl( \F - \F^{-T}
           + \frac{\alpha}{\mu} \, \F \ba \otimes \ba \Bigr),
\end{equation*}
which shows that, in absence of external loads and with constant $\mu$, we are as a matter of fact estimating the ratio between $\alpha$ and $\mu$. We can even claim that the knowledge of the value of the elastic modulus is immaterial because we estimate the ratio, which is in fact informative of the presence of a scar (lower contractility and/or higher stiffness.)
This basic observation highlights that $\mu$ and $\alpha$ can be estimated independently only when longitudinal data are available, e.g., the displacement field over the entire cardiac cycle.

More generally, if $\mu(\bX)$ is non-constant and known, we can recover $\alpha$ with good accuracy; otherwise, $\alpha$ will implicitly compensate the reduced stiffness. This is exemplified in Fig.~\ref{fig:robustness2noise} (left panel). Interestingly, the scar can still be detected with good accuracy.

\subsection{Robustness to noise and subsampling}
\label{sec:noise}

\begin{figure}[t]
	\centering
	\includegraphics[width=0.34\textwidth,trim=10 0 40 40,clip]{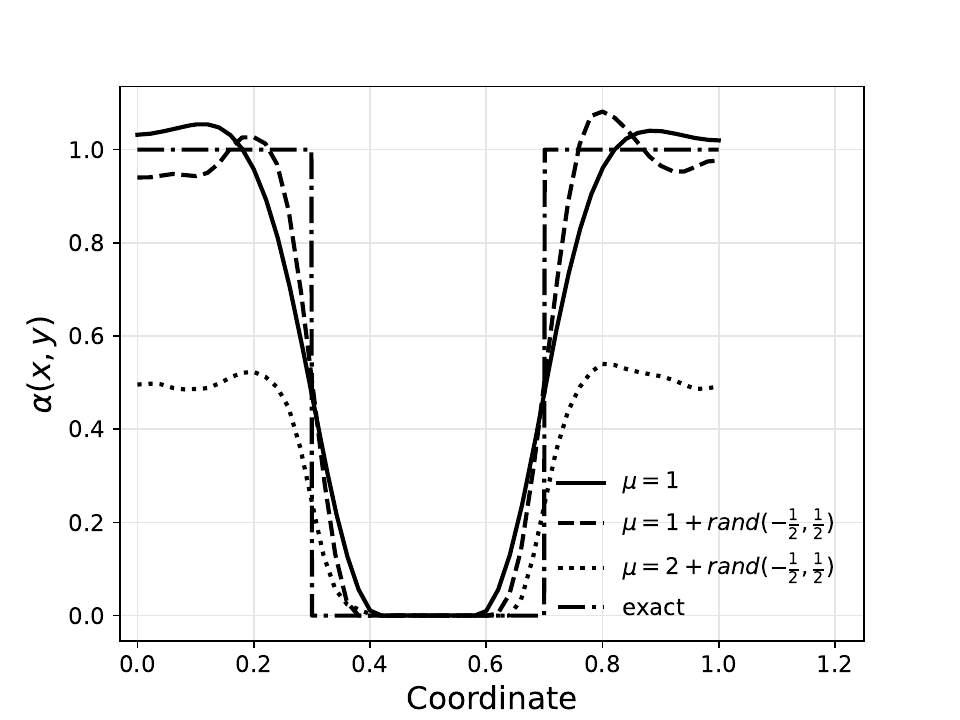} \hfill
    \includegraphics[width=0.25\textwidth]{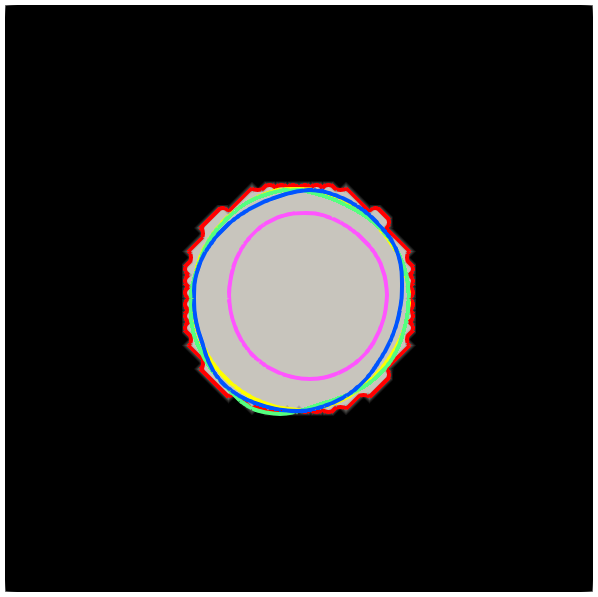} \hfill
	\includegraphics[width=0.34\textwidth,trim=10 0 40 40,clip]{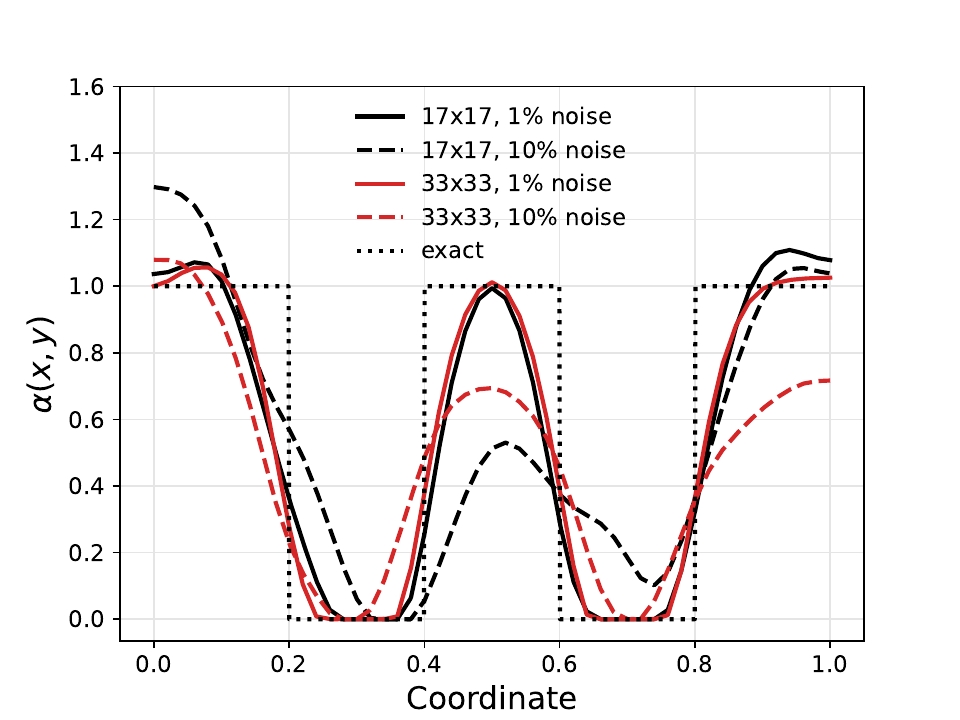}

 \caption{
\rev{(Left) Robustness to an incorrect choice of the forward model. The ground truth is based on various values of $\mu$, including noise. In the inverse problem, we assume $\mu=1$. The profiles are plotted along the direction orthogonal to the fiber orientation.
(Center) Effect of high, uncorrelated noise perturbing the ground truth: 1\% (yellow), 5\% (green), 10\% (blue), and 20\% (purple). Isolines for $\alpha = 0.4$.
(Right) Effect of subsampling the ground truth on a coarser mesh.
All the results are with full observation operator and $\mRH$ regularization.}}
\label{fig:robustness2noise}
\end{figure}

\linelabel{ll:noisy_gt}
In the previous examples we considered a noise level of about $1\%$, which is low for real applications but is reasonable for testing the numerical stability of the inverse problem. 
In Fig.~\ref{fig:robustness2noise} (middle panel) we repeated the experiment with a level of uncorrelated noise applied to the ground truth solution up to $20\%$. 
For each simulation, to set the optimal regularization weight $\lambda_\text{opt}$, the L-curve has been recalculated.
We notice that the reconstruction is still accurate for a noise level below $10\%$, while for a $20\%$ noise the scar is detected in the right position but with a smaller size.
 
\linelabel{ll:coarse_gt}
Furthermore, we tested the effect of using a coarser resolution for the ground truth, while maintaining a higher resolution for the solution of the inverse problem. Specifically, we generated the ground truth at high resolution, and then we subsampled it on either a $17\times 17$ or $33\times 33$ uniform grid. Additionally, we included a $1\%$ or $10\%$ noise in the simulations. The ground truth profile for the scar was the same as in Fig.~\ref{tab::shape}, described later. Results, collected in Fig.~\ref{fig:robustness2noise} (right panel), show a good accuracy in detecting the size of the scar and the presence of channels, except for the $17\times 17$ case with $10\%$ noise, where the channels are asymmetric and of smaller size.
}

\rev{
\subsection{Incompressibility}
}
In this part, we test the effect of the incompressibility assumption in the forward model. We use the formulation in Eq.~\eqref{eq::P_p_inc}. The incompressibility constraint is enforced for both the ground truth model and the inverse problem. \linelabel{ll:incompress}\rev{Concerning the discretization, we solved the for the mixed formulation of the forward model with $\mathbb{P}^2-\mathbb{P}^1$ Taylor-Hood finite elements (piecewise quadratic elements for the displacement, linear for the pressure). The adjoint model also resulted in a (linear) elasticity problem with a linearized incompressibility constraint, and it was solved with the same Taylor-Hood elements of the forward model.}

\begin{figure}
	\centering
	\begin{tabular}{ccc}
		$\mRL$ & $\mRTV$ & $\mRH$ \\
		\includegraphics[width=0.3\textwidth,trim=10 0 45 40,clip]{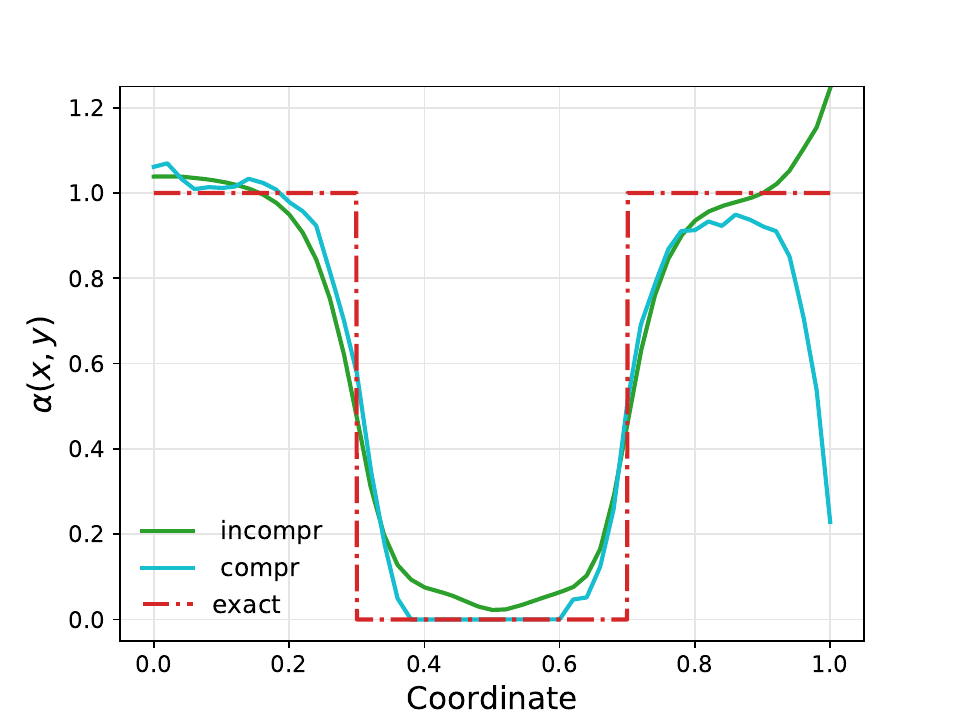} &
		\includegraphics[width=0.3\textwidth,trim=10 0 45 40,clip]{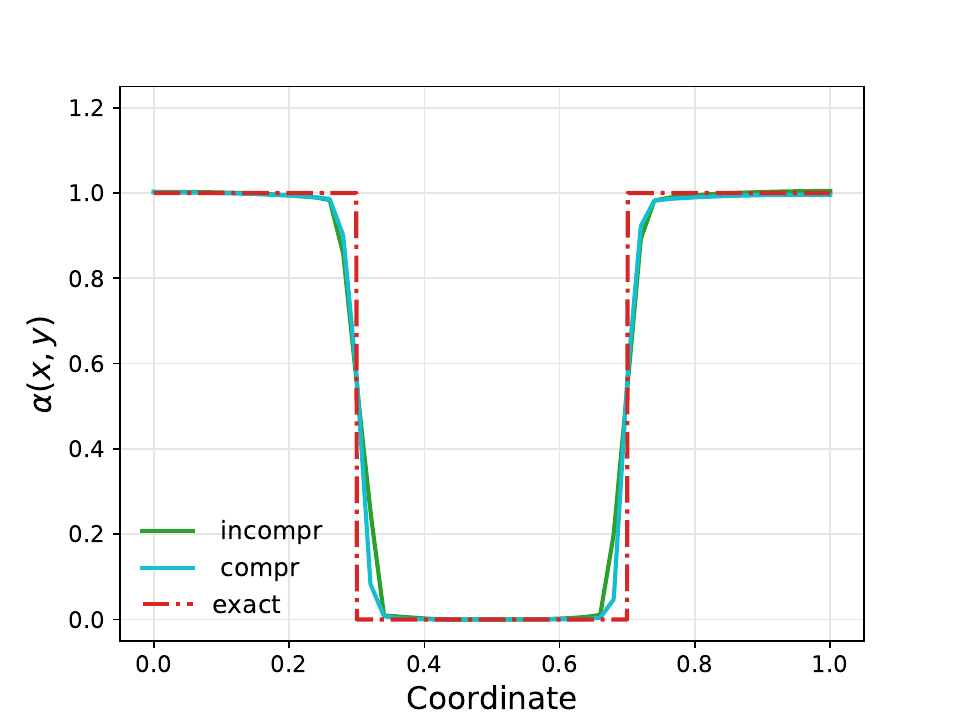} &
		\includegraphics[width=0.3\textwidth,trim=10 0 45 40,clip]{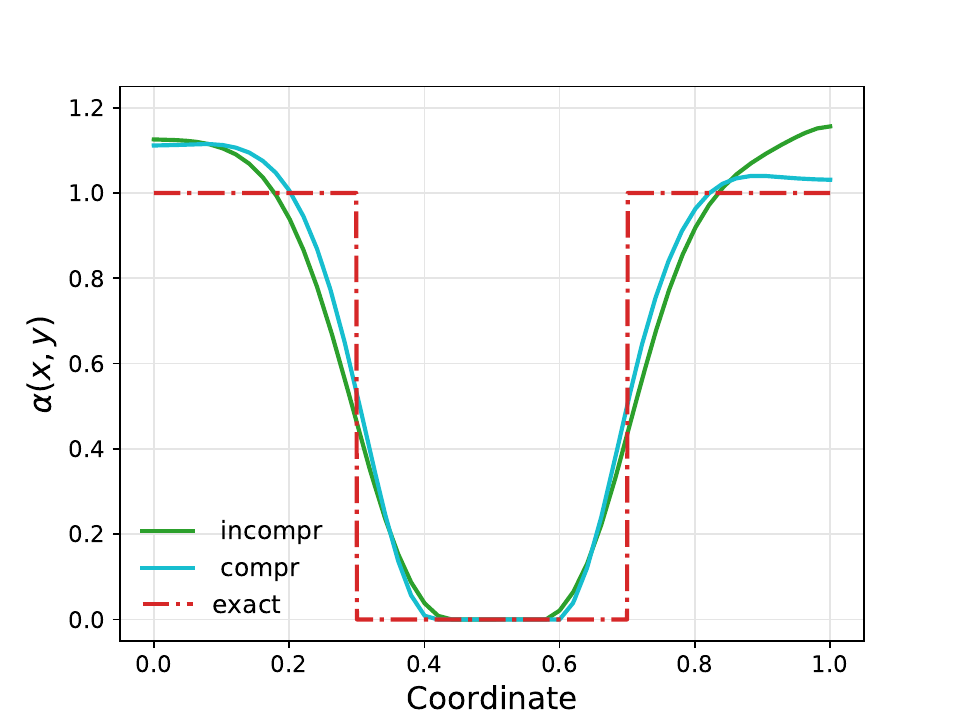}
		
	\end{tabular}
	\caption{Effect of the incompressibility hypothesis on contractility reconstruction. Comparison between the exact profile of $\alpha(x,y)$ (red) and the estimated contractility profiles for the fully incompressible material model (green) and the compressible material model (cyan) along the direction parallel to the fiber orientation. 
    For each chosen regularization we set $\lambda = \lambda_\text{opt}$.}
	\label{fig::incompressibility}
\end{figure}

Results are shown in Fig.~\ref{fig::incompressibility}.
The incompressibility partially reduces the observability issue related to the choice of $\mRL$ in the contractility estimation; however, it overestimates $\alpha(x,y)$ on the free boundary orthogonal to the fiber direction.
Conversely, the reconstruction obtained with the incompressible model fully match the ground truth, when choosing  $\mRTV$ as regularization. In this case, we also observe a complete agreement between the compressible and the incompressible material model, which can be interpreted as a symptom of over-regularization.
For $\mRH$, the compressible model performs better than the incompressible model with reference to the ground truth since, as for the $\mRL$, the latter tends to overestimate the contractility at the free boundary.

\subsection{Boundary observation}
\label{sec:bcobs}

\begin{figure}[t]
	\centering
	\begin{tabular}{ccc}
	$\mRL$ & $\mRTV$ & $\mRH$ \\
	\includegraphics[width=0.27\textwidth]{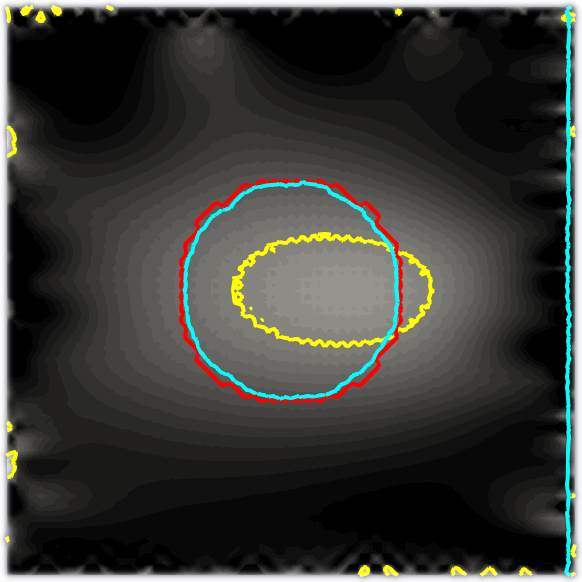} &
	\includegraphics[width=0.27\textwidth]{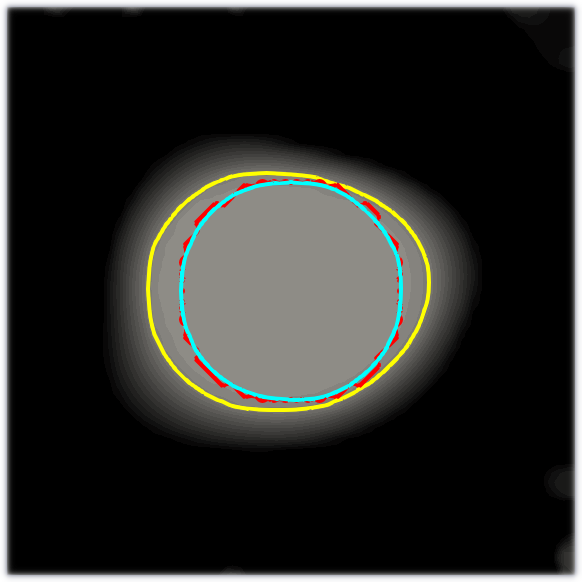} & 
	\includegraphics[width=0.27\textwidth]{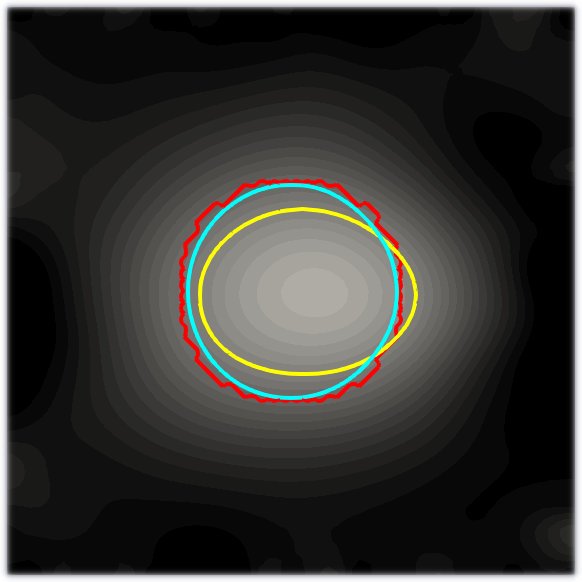} \\
    \multicolumn{2}{c}{\raisebox{1em}{\includegraphics[width=0.3\linewidth]{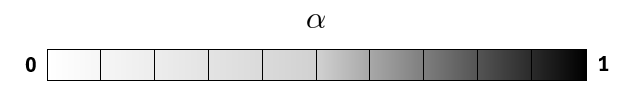}}} & \includegraphics[width=0.32\textwidth]{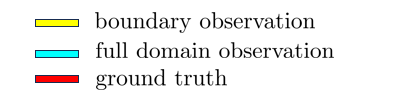} 
	\end{tabular}
	\caption{Effect of restricting observations to the boundary. Scar profiles obtained from the minimization of the cost functional \eqref{eq::functional_bd} with the boundary observation operator for all the possible regularizing term\rev{s}: $\mRL$ (left), $\mRTV$ (center), and $\mRH$ (right). Isolines $\alpha = 0.4$ for the boundary observation (yellow), for the full domain observation (blue) and for the ground truth (red) are superimposed to the contractility pattern. In each case we set $\lambda = \lambda_\text{opt}$. The value of $\alpha$ was rescaled to the range $[0,1]$ for visual clarity.}
	\label{fig:bd_obs}
\end{figure}

We now address a particular case, considering a restriction of the data fidelity term to the boundary of the domain, so that the cost functional reads as follows
\begin{equation}
	\mathcal{J}(\bu,\alpha) = \frac{1}{2} \int_{\rev{\GammaN}} | \bu - \bu_{o}|^2 \:\dd S 
	+ \lambda\mR(\alpha).
\label{eq::functional_bd}
\end{equation}
\rev{The corresponding adjoint problem is Eq.~\eqref{eq::adjointBC}.}

Fig.~\ref{fig:bd_obs} shows the estimated contractility function $\alpha(x,y)$ for each possible choice for the regularizing term.
As expected, in all the three cases the scar profile (yellow contours) is captured with a significantly lower level of accuracy with respect to the case where the observation operator is extended to the whole domain (blue contours). 
More precisely, while the choice of $\mRL$ and $\mRH$ as regularization leads to an underestimation of the scar area, the choice of  $\mRTV$ results in an opposite behavior. 
Overall, the scar position is properly identified. However, the shape appears excessively elongated in the direction of the fibers, see Fig \ref{fig:bd_obs} ($\mRL$ and $\mRH$).


\subsection{Robustness to scar shape variations}

\begin{figure}[htp]
	\centering
	\begin{tabular}{c c c}
	 ground truth  & full observation & boundary observation\\

	    \includegraphics[width=0.24\linewidth]{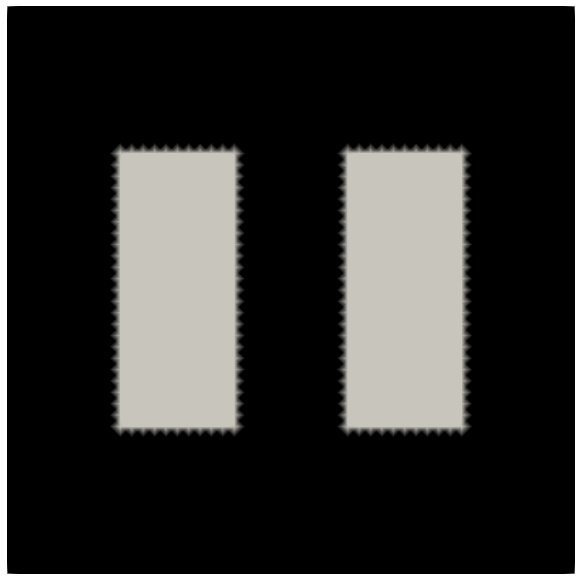} & \includegraphics[width=0.24\linewidth]{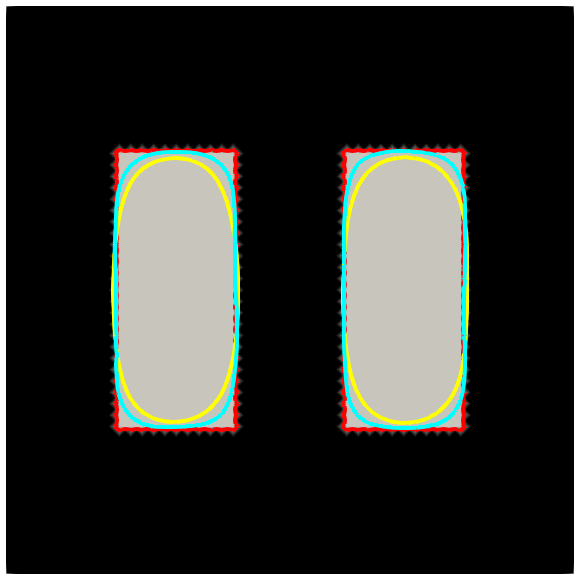} & \includegraphics[width=0.24\linewidth]{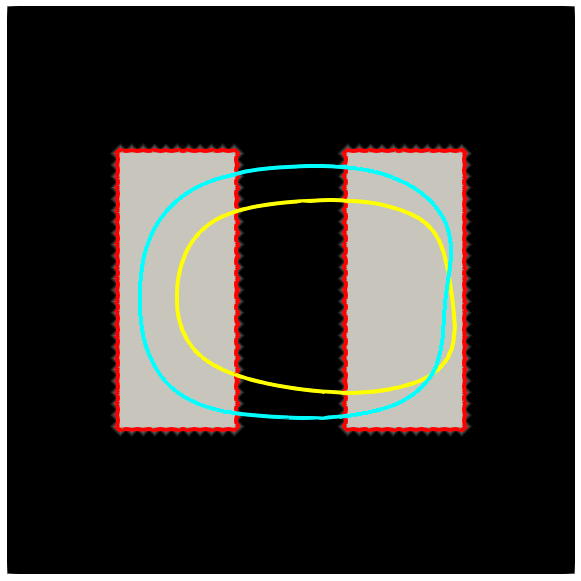}\\
		  \includegraphics[width=0.24\linewidth]{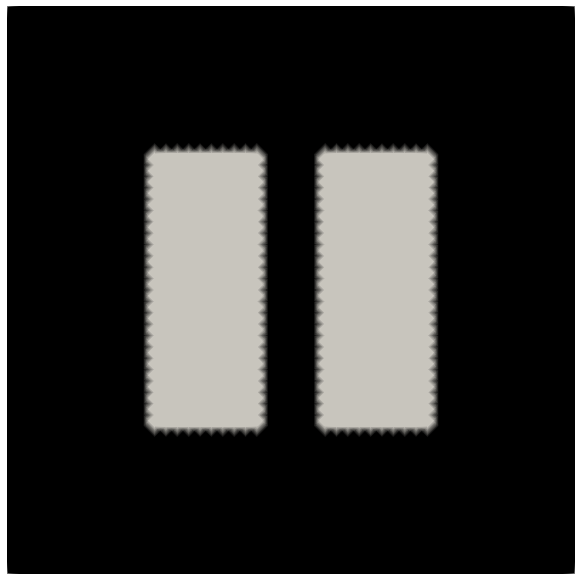} & \includegraphics[width=0.24\linewidth]{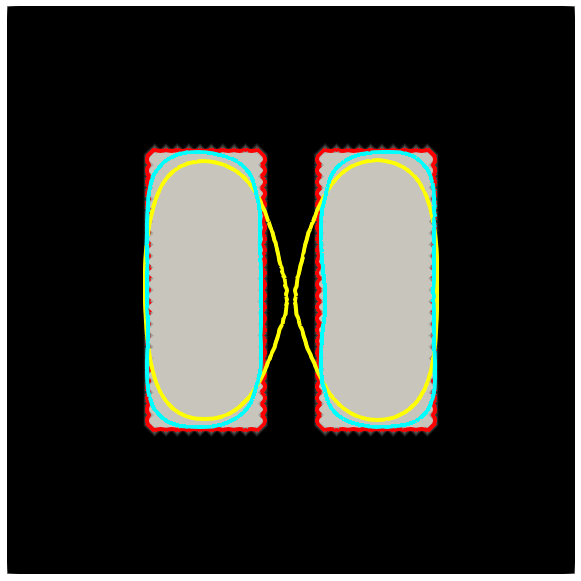} & \includegraphics[width=0.24\linewidth]{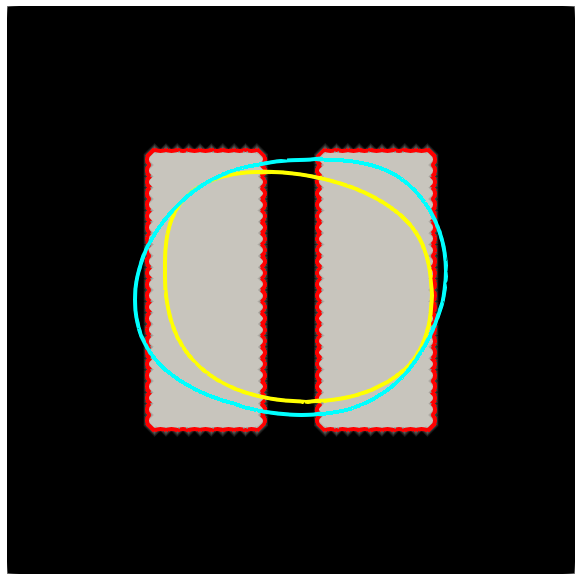}\\
        \includegraphics[width=0.24\linewidth]{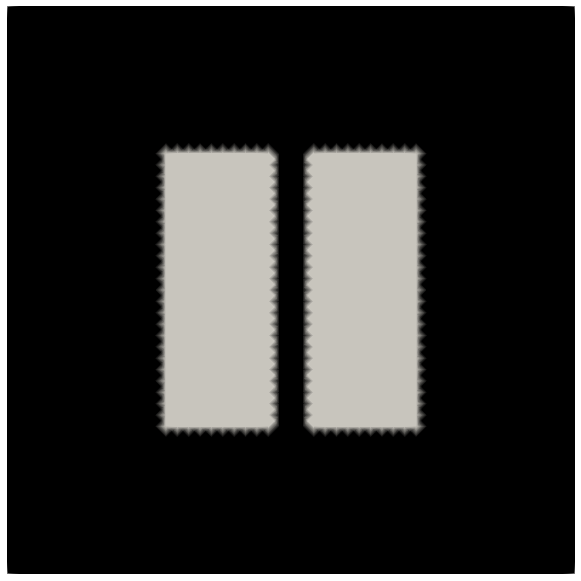} & \includegraphics[width=0.24\linewidth]{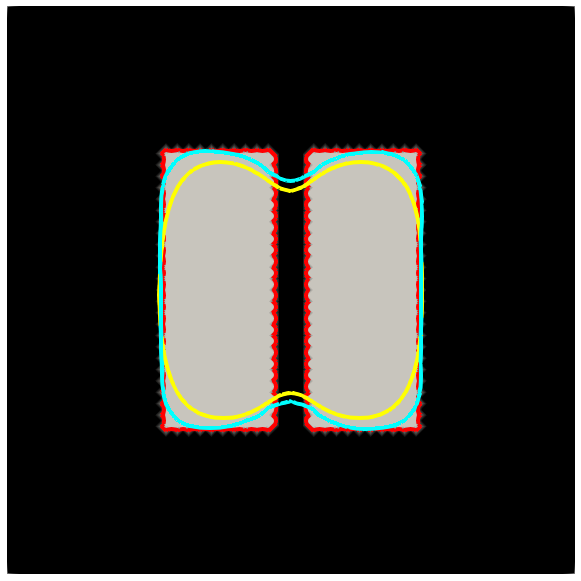} & \includegraphics[width=0.24\linewidth]{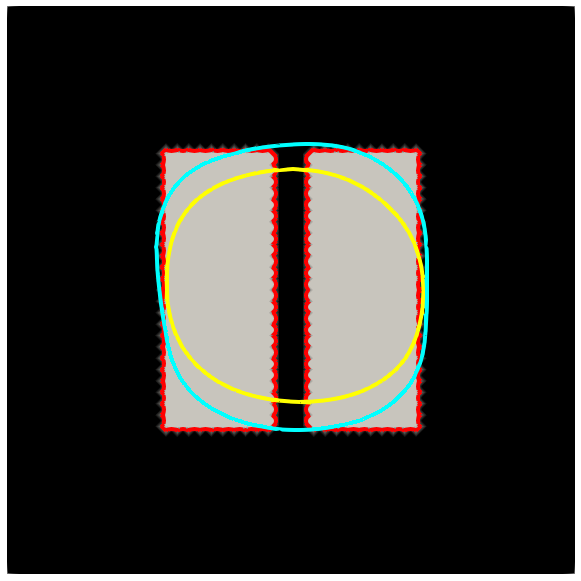}\\
%
	\multicolumn{2}{c}{\includegraphics[width=0.3\linewidth]   {img/legenda_alpha_oriz.PNG}} &
        \includegraphics[width=0.26\linewidth]{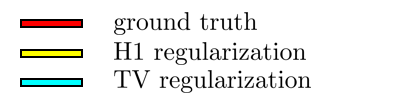}
       \\
	\end{tabular}
	\caption{Sensitivity to the scar shape. The accuracy of the algorithm in reconstructing the contractility profile has been tested for three different scar conformations, obtained by progressively reducing the distance between two rectangular scars (Left: from top to bottom row). The test was carried out both by adopting the observation operator extended to the whole domain (center) and the one restricted to the boundary (right). Isolines $\alpha(x,y)$ = \rev{0.4} for the ground truth (red), the H1 regularization (yellow) and the TV regularization (blue). \rev{The ground truth is shown in the background.} For each chosen regularization we set $\lambda = \lambda_\text{opt}$.
    }
	\label{tab::shape}
\end{figure}

The results obtained by varying the scar shape are shown in Fig.~\ref{tab::shape}. Motivated by clinical observations, the ground truth scar geometry consist of two rectangles separated by a narrow strip of healthy tissue.
We notice that the accuracy of the reconstructed contractility $\alpha(x,y)$ decreases as the distance between the two scars diminishes. Overall, the use of the observation operator restricted to the boundary of the domain dramatically affects the quality of the model prediction in all the proposed test cases.
On the contrary, the performance of the algorithm is significantly improved,  except in the most extreme case (Fig.~\ref{tab::shape}, bottom row), by using the full domain observation combined with TV regularization. This choice allows for correct and accurate identification of the areas where contractility is impaired.

\section{Discussion and conclusions}
\label{sec:conclusions}


Determining the contractility of the cardiac tissue from displacement measurements is an important topic in cardiac modeling and patient-specific applications~\cite{asner,finsberg,kovacheva}. As other mechanical and electrical parameters, the local contractility cannot be directly measured \emph{in vivo}, and must be deduced indirectly. Scarred and ischemic regions, where contractility is low or absent, can be delineated from low-voltage regions or from late enhancement MRI, for instance. Here, \rev{similar to quasi-static elastography~\cite{doyley2012model}}, we have focused on kinematic information. Myocardial motion and strain can be acquired non-invasively by standard imaging techniques such as cardiac MRI or echo, or in an invasive manner with electroanatomical mapping systems~\cite{Maffessanti2020scar}. Importantly, cardiac motion can be modelled by a balance of forces that includes active stress, offering the opportunity for the determination of contractility through an inverse problem solution.

After stating a suitable cost functional, designed to minimize the discrepancy between the observed displacement and the one predicted by the model, we have straightforwardly derived the adjoint equation, the state equation, and the optimality condition for the reconstruction of the contractility function. Given the ill-posedness of the inverse problem, a regularization term has to be introduced. We have then tackled the solution of the problem from the numerical perspective as an optimization problem, thus adopting the L-BFGS-B algorithm to minimize the cost functional subjected to the force balance equation and its boundary conditions.
\linelabel{ll:discussion_severe}\rev{We have numerically tested our approach in a synthetic case with a region of impaired contractility sharply separated from the active one, and with 1\% of noise superimposed to the observed displacements. Within this framework, we have explored the stability of our approach in the presence of model perturbations, such as an imperfect knowledge of the forward problem or the ground truth. We also tested three different regularizing terms discussing the prod and cons of each of them.}

The numerical results show that the proposed approach is very effective and the H1 regularization is a good compromise both in terms of accuracy and efficiency: it overcomes both the limitation related to the slow convergence of the approximated TV regularization and the observability issues associated to the L2 regularization. The apparent superior ability of the TV regularization in capturing sharp boundaries (see Fig.~\ref{fig::incompressibility}) should not be overemphasized: while we have purposely designed challenging numerical tests for the inversion method, in practical applications infarcted regions typically exhibit a smooth transition from healthy to dead cells. \rev{Conversely, if the pattern of inactive region is fragmented like in Fig.~\ref{tab::shape}, the increased computational cost of TV is well justified by the improved accuracy in detecting a thin active region.}

Overall, the predicted local contractility is in good agreement with the ground truth, showing robustness of the results with respect to random noise both on the data and on the local value of the elastic modulus $\mu$ and the fibers direction $\theta$. Moreover, when the TV and H1 regularizations are adopted, the proposed method appears to be able to predict with a good level of accuracy the local contractility even when the observed data are restricted to the boundary. \rev{The inverse problem with boundary observation is conceptually similar to the Calderón problem in electrical impedance tomography~\cite{cheney1999electrical}, which aims to delineate the organs from sharp transitions in the electrical impedance of the body, reconstructed from boundary measurements.} This result is of particular interest with the perspective to tackle three dimensional cardiac data when the displacement is measured on \rev{(part of) the cardiac surface} only. 

Some methods proposed in the literature allow for the estimation of contractility as a region-wise constant function \cite{chabiniok2012estimation, finsberg, Sjur2015Stacom}. The left ventricular wall is divided into 17 areas, known as AHA segments, and a constant contractility value is assigned to each of these segments by solving a discrete inverse problem, which exploits as datum the displacement acquired by means of imaging techniques of one or more points belonging to each AHA segment.
\rev{Being low-dimensional, region-wise approaches are computationally attractive and deemed to surrogate modeling. However, these methods do not scale well with the number of parameters, thus they cannot clearly delineate the scar with higher resolution.}
Conversely, our \rev{adjoint-based} approach enables an estimation of the local contractility as a continuous function across the entire domain, thus accurately identifying the scar profile without compromising the computational costs.

More recently, \citet{kovacheva} addressed the problem of estimating the contractility and scarred tissue is 3D real geometries over an entire cardiac cycle. Their method is based on an inverse problem where the tangent map from the contractility to the displacement is estimated via finite differences. At numerical level, this matrix is dense and possibly large, thus the method may suffer of scalability issues. The adjoint approach proposed in this work overcomes this problem, because we can compute the action of the tangent map as the solution of a linear problem. Secondly, we are able to recover the contractility pattern in a large strain regime and for general hyperelastic materials. Interestingly, we did not observe a large sensitivity to fiber orientation for the detection of the scar, but this aspect should be further investigated in more a complex geometrical setup.

The method proposed in this paper suffers of course from some limitations. First, from a methodological point of view, the method might converge towards a local minimum of the cost functional, while the convergence to the global minimum is not guaranteed. \rev{Running several initial conditions, we have never observed such a behavior in our experiments.
In the same vein, for large strain the forward problem might have multiple equilibria; however these ranges would be very exotic in the case in the context of cardiac mechanics~\cite{ambrosi2012active}.
}
Second, we have tested the approach only with synthetic data and simplified modelling assumption; in particular, we have adopted a simple strain energy that cannot be naively applied to cardiac modeling. In this respect, our methodology can be straightforwardly modified to incorporate more appropriate material laws~\cite{guccione1991passive, holzapfel2009constitutive}.
\rev{
Our preliminary experiments in this direction are encouraging, since we observed no degradation of performance when we employed the incompressible Holzapfel-Ogden constitutive law~\cite{holzapfel2009constitutive}.
}
In view of the incorporation of clinical data, we plan to further develop this work to 3D geometries. In the same vein, this methodology could be easily reformulated to determine the orientation of the myocardial fibers on the basis of kinematic data in an healthy heart~\cite{Lubrecht2021PIEMAP,RuizPINN2022}. 

In conclusion, our method provides a good framework for the identification of mechanical parameters of cardiac models from non-invasive clinical data, and it can significantly improve the definition of patient-specific models for precision cardiology.
Finally, we remark that the validity of the proposed approach is not limited to cardiac electromechanics but the proposed method can be used for estimating the local contractility in a generic active material, whether natural or artificial. \rev{Our work might even be recast in a control framework; for very large strain it might be seen as an approach to produce target shapes on the basis of activation or material rearrangement.}

\section*{Declaration of competing interest}
The authors declare that they have no known competing financial interests or personal relationships that could have appeared to influence the work reported in this paper.

\section*{Acknowledgments}
DA and GP were supported by the PRIN Research Project 2020F3NCPX. DA and GP acknowledges the support of INDAM-GNFM. GP was partially supported by INDAM-GNFM \textit{``Fondo Progetto Giovani 2023''}. SP was supported by the SNSF project ``CardioTwin'' (no.~214817), the CSCS grant no.~s1074, and the PRIN-PNRR project no.~P2022N5ZNP. SP acknowledges the support of INDAM-GNCS.

\bibliographystyle{elsarticle-harv} 
\bibliography{References}

\begin{thebibliography}{52}
\expandafter\ifx\csname natexlab\endcsname\relax\def\natexlab#1{#1}\fi
\providecommand{\url}[1]{\texttt{#1}}
\providecommand{\href}[2]{#2}
\providecommand{\path}[1]{#1}
\providecommand{\DOIprefix}{doi:}
\providecommand{\ArXivprefix}{arXiv:}
\providecommand{\URLprefix}{URL: }
\providecommand{\Pubmedprefix}{pmid:}
\providecommand{\doi}[1]{\href{http://dx.doi.org/#1}{\path{#1}}}
\providecommand{\Pubmed}[1]{\href{pmid:#1}{\path{#1}}}
\providecommand{\bibinfo}[2]{#2}
\ifx\xfnm\relax \def\xfnm[#1]{\unskip,\space#1}\fi
\bibitem[{Aln{\ae}s et~al.(2014)Aln{\ae}s, Logg, {\O}lgaard, Rognes and
  Wells}]{alnaes2014unified}
\bibinfo{author}{Aln{\ae}s, M.}, \bibinfo{author}{Logg, A.},
  \bibinfo{author}{{\O}lgaard, K.}, \bibinfo{author}{Rognes, M.},
  \bibinfo{author}{Wells, G.}, \bibinfo{year}{2014}.
\newblock \bibinfo{title}{Unified form language: A domain-specific language for
  weak formulations of partial differential equations}.
\newblock \bibinfo{journal}{ACM Transactions on Mathematical Software (TOMS)}
  \bibinfo{volume}{40}, \bibinfo{pages}{1--37}.
\bibitem[{Ambrosi and Pezzuto(2012)}]{ambrosi2012active}
\bibinfo{author}{Ambrosi, D.}, \bibinfo{author}{Pezzuto, S.},
  \bibinfo{year}{2012}.
\newblock \bibinfo{title}{Active stress vs. active strain in mechanobiology:
  constitutive issues}.
\newblock \bibinfo{journal}{Journal of Elasticity} \bibinfo{volume}{107},
  \bibinfo{pages}{199--212}.
\bibitem[{Amestoy et~al.(2001)Amestoy, Duff, Koster and L'Excellent}]{MUMPS:1}
\bibinfo{author}{Amestoy, P.}, \bibinfo{author}{Duff, I.S.},
  \bibinfo{author}{Koster, J.}, \bibinfo{author}{L'Excellent, J.Y.},
  \bibinfo{year}{2001}.
\newblock \bibinfo{title}{A fully asynchronous multifrontal solver using
  distributed dynamic scheduling}.
\newblock \bibinfo{journal}{SIAM Journal on Matrix Analysis and Applications}
  \bibinfo{volume}{23}, \bibinfo{pages}{15--41}.
\bibitem[{Asner et~al.(2016)Asner, Hadjicharalambous, Chabiniok, Peresutti,
  Sammut, Wong, Carr-White, Chowienczyk, Lee, King et~al.}]{asner}
\bibinfo{author}{Asner, L.}, \bibinfo{author}{Hadjicharalambous, M.},
  \bibinfo{author}{Chabiniok, R.}, \bibinfo{author}{Peresutti, D.},
  \bibinfo{author}{Sammut, E.}, \bibinfo{author}{Wong, J.},
  \bibinfo{author}{Carr-White, G.}, \bibinfo{author}{Chowienczyk, P.},
  \bibinfo{author}{Lee, J.}, \bibinfo{author}{King, A.}, et~al.,
  \bibinfo{year}{2016}.
\newblock \bibinfo{title}{Estimation of passive and active properties in the
  human heart using {3D} tagged {MRI}}.
\newblock \bibinfo{journal}{Biomechanics and modeling in mechanobiology}
  \bibinfo{volume}{15}, \bibinfo{pages}{1121--1139}.
\bibitem[{Augenstein et~al.(2005)Augenstein, Cowan, LeGrice, Nielsen and
  Young}]{augenstein2005method}
\bibinfo{author}{Augenstein, K.}, \bibinfo{author}{Cowan, B.},
  \bibinfo{author}{LeGrice, I.}, \bibinfo{author}{Nielsen, P.},
  \bibinfo{author}{Young, A.}, \bibinfo{year}{2005}.
\newblock \bibinfo{title}{Method and apparatus for soft tissue material
  parameter estimation using tissue tagged magnetic resonance imaging}.
\newblock \bibinfo{journal}{J. Biomech. Eng.} \bibinfo{volume}{127},
  \bibinfo{pages}{148--157}.
\bibitem[{Auricchio et~al.(2013)Auricchio, Beir{\~a}o~da Veiga, Lovadina,
  Reali, Taylor and Wriggers}]{auricchio2013approximation}
\bibinfo{author}{Auricchio, F.}, \bibinfo{author}{Beir{\~a}o~da Veiga, L.},
  \bibinfo{author}{Lovadina, C.}, \bibinfo{author}{Reali, A.},
  \bibinfo{author}{Taylor, R.L.}, \bibinfo{author}{Wriggers, P.},
  \bibinfo{year}{2013}.
\newblock \bibinfo{title}{Approximation of incompressible large deformation
  elastic problems: some unresolved issues}.
\newblock \bibinfo{journal}{Computational Mechanics} \bibinfo{volume}{52},
  \bibinfo{pages}{1153--1167}.
\bibitem[{Balay et~al.(2023)Balay, Abhyankar, Adams, Benson, Brown, Brune,
  Buschelman, Constantinescu, Dalcin, Dener, Eijkhout, Faibussowitsch, Gropp,
  Hapla, Isaac, Jolivet, Karpeev, Kaushik, Knepley, Kong, Kruger, May, McInnes,
  Mills, Mitchell, Munson, Roman, Rupp, Sanan, Sarich, Smith, Zampini, Zhang,
  Zhang and Zhang}]{petsc-web-page}
\bibinfo{author}{Balay, S.}, \bibinfo{author}{Abhyankar, S.},
  \bibinfo{author}{Adams, M.F.}, \bibinfo{author}{Benson, S.},
  \bibinfo{author}{Brown, J.}, \bibinfo{author}{Brune, P.},
  \bibinfo{author}{Buschelman, K.}, \bibinfo{author}{Constantinescu, E.M.},
  \bibinfo{author}{Dalcin, L.}, \bibinfo{author}{Dener, A.},
  \bibinfo{author}{Eijkhout, V.}, \bibinfo{author}{Faibussowitsch, J.},
  \bibinfo{author}{Gropp, W.D.}, \bibinfo{author}{Hapla, V.},
  \bibinfo{author}{Isaac, T.}, \bibinfo{author}{Jolivet, P.},
  \bibinfo{author}{Karpeev, D.}, \bibinfo{author}{Kaushik, D.},
  \bibinfo{author}{Knepley, M.G.}, \bibinfo{author}{Kong, F.},
  \bibinfo{author}{Kruger, S.}, \bibinfo{author}{May, D.A.},
  \bibinfo{author}{McInnes, L.C.}, \bibinfo{author}{Mills, R.T.},
  \bibinfo{author}{Mitchell, L.}, \bibinfo{author}{Munson, T.},
  \bibinfo{author}{Roman, J.E.}, \bibinfo{author}{Rupp, K.},
  \bibinfo{author}{Sanan, P.}, \bibinfo{author}{Sarich, J.},
  \bibinfo{author}{Smith, B.F.}, \bibinfo{author}{Zampini, S.},
  \bibinfo{author}{Zhang, H.}, \bibinfo{author}{Zhang, H.},
  \bibinfo{author}{Zhang, J.}, \bibinfo{year}{2023}.
\newblock \bibinfo{title}{{PETS}c {W}eb page}.
\newblock \bibinfo{howpublished}{\url{https://petsc.org/}}.
\newblock \URLprefix \url{https://petsc.org/}.
\bibitem[{Balay et~al.(1997)Balay, Gropp, McInnes and Smith}]{petsc-efficient}
\bibinfo{author}{Balay, S.}, \bibinfo{author}{Gropp, W.D.},
  \bibinfo{author}{McInnes, L.C.}, \bibinfo{author}{Smith, B.F.},
  \bibinfo{year}{1997}.
\newblock \bibinfo{title}{Efficient management of parallelism in object
  oriented numerical software libraries}, in: \bibinfo{editor}{Arge, E.},
  \bibinfo{editor}{Bruaset, A.M.}, \bibinfo{editor}{Langtangen, H.P.} (Eds.),
  \bibinfo{booktitle}{Modern Software Tools in Scientific Computing},
  \bibinfo{publisher}{Birkh{\"{a}}user Press}. pp. \bibinfo{pages}{163--202}.
\bibitem[{Ball(1976)}]{ball1976convexity}
\bibinfo{author}{Ball, J.}, \bibinfo{year}{1976}.
\newblock \bibinfo{title}{Convexity conditions and existence theorems in
  nonlinear elasticity}.
\newblock \bibinfo{journal}{Archive for rational mechanics and Analysis}
  \bibinfo{volume}{63}, \bibinfo{pages}{337--403}.
\bibitem[{Budd and Nichols(2011)}]{budd2011regularization}
\bibinfo{author}{Budd, C.}, \bibinfo{author}{Nichols, N.},
  \bibinfo{year}{2011}.
\newblock \bibinfo{title}{Regularization techniques for ill-posed inverse
  problems in data assimilation}.
\newblock \bibinfo{journal}{Computers \& fluids} \bibinfo{volume}{46},
  \bibinfo{pages}{168--173}.
\bibitem[{Byrd et~al.(1995)Byrd, Lu, Nocedal and Zhu}]{byrd1995limited}
\bibinfo{author}{Byrd, R.H.}, \bibinfo{author}{Lu, P.},
  \bibinfo{author}{Nocedal, J.}, \bibinfo{author}{Zhu, C.},
  \bibinfo{year}{1995}.
\newblock \bibinfo{title}{A limited memory algorithm for bound constrained
  optimization}.
\newblock \bibinfo{journal}{SIAM Journal on scientific computing}
  \bibinfo{volume}{16}, \bibinfo{pages}{1190--1208}.
\bibitem[{Chabiniok et~al.(2012)Chabiniok, Moireau, Lesault, Rahmouni, Deux and
  Chapelle}]{chabiniok2012estimation}
\bibinfo{author}{Chabiniok, R.}, \bibinfo{author}{Moireau, P.},
  \bibinfo{author}{Lesault, P.}, \bibinfo{author}{Rahmouni, A.},
  \bibinfo{author}{Deux, J.}, \bibinfo{author}{Chapelle, D.},
  \bibinfo{year}{2012}.
\newblock \bibinfo{title}{Estimation of tissue contractility from cardiac
  cine-mri using a biomechanical heart model}.
\newblock \bibinfo{journal}{Biomechanics and modeling in mechanobiology}
  \bibinfo{volume}{11}, \bibinfo{pages}{609--630}.
\bibitem[{Chan et~al.(2005)Chan, Esedoglu, Park, Yip et~al.}]{chan2005recent}
\bibinfo{author}{Chan, T.}, \bibinfo{author}{Esedoglu, S.},
  \bibinfo{author}{Park, F.}, \bibinfo{author}{Yip, A.}, et~al.,
  \bibinfo{year}{2005}.
\newblock \bibinfo{title}{Recent developments in total variation image
  restoration}.
\newblock \bibinfo{journal}{Mathematical Models of Computer Vision}
  \bibinfo{volume}{17}, \bibinfo{pages}{17--31}.
\bibitem[{Cheney et~al.(1999)Cheney, Isaacson and
  Newell}]{cheney1999electrical}
\bibinfo{author}{Cheney, M.}, \bibinfo{author}{Isaacson, D.},
  \bibinfo{author}{Newell, J.C.}, \bibinfo{year}{1999}.
\newblock \bibinfo{title}{Electrical impedance tomography}.
\newblock \bibinfo{journal}{SIAM review} \bibinfo{volume}{41},
  \bibinfo{pages}{85--101}.
\bibitem[{Chung et~al.(2021)Chung, Parsons and Zheng}]{chung2021magnetically}
\bibinfo{author}{Chung, H.J.}, \bibinfo{author}{Parsons, A.},
  \bibinfo{author}{Zheng, L.}, \bibinfo{year}{2021}.
\newblock \bibinfo{title}{Magnetically controlled soft robotics utilizing
  elastomers and gels in actuation: A review}.
\newblock \bibinfo{journal}{Advanced Intelligent Systems} \bibinfo{volume}{3},
  \bibinfo{pages}{2000186}.
\bibitem[{Cicci et~al.(2023)Cicci, Fresca, Guo, Manzoni and
  Zunino}]{cicci2023uncertainty}
\bibinfo{author}{Cicci, L.}, \bibinfo{author}{Fresca, S.},
  \bibinfo{author}{Guo, M.}, \bibinfo{author}{Manzoni, A.},
  \bibinfo{author}{Zunino, P.}, \bibinfo{year}{2023}.
\newblock \bibinfo{title}{Uncertainty quantification for nonlinear solid
  mechanics using reduced order models with gaussian process regression}.
\newblock \bibinfo{journal}{Computers \& Mathematics with Applications}
  \bibinfo{volume}{149}, \bibinfo{pages}{1--23}.
\bibitem[{Delingette et~al.(2011)Delingette, Billet, Wong, Sermesant, Rhode,
  Ginks, Rinaldi, Razavi and Ayache}]{delingette2011personalization}
\bibinfo{author}{Delingette, H.}, \bibinfo{author}{Billet, F.},
  \bibinfo{author}{Wong, K.}, \bibinfo{author}{Sermesant, M.},
  \bibinfo{author}{Rhode, K.}, \bibinfo{author}{Ginks, M.},
  \bibinfo{author}{Rinaldi, C.}, \bibinfo{author}{Razavi, R.},
  \bibinfo{author}{Ayache, N.}, \bibinfo{year}{2011}.
\newblock \bibinfo{title}{Personalization of cardiac motion and contractility
  from images using variational data assimilation}.
\newblock \bibinfo{journal}{IEEE transactions on biomedical engineering}
  \bibinfo{volume}{59}, \bibinfo{pages}{20--24}.
\bibitem[{Doyley(2012)}]{doyley2012model}
\bibinfo{author}{Doyley, M.M.}, \bibinfo{year}{2012}.
\newblock \bibinfo{title}{Model-based elastography: a survey of approaches to
  the inverse elasticity problem}.
\newblock \bibinfo{journal}{Physics in Medicine \& Biology}
  \bibinfo{volume}{57}, \bibinfo{pages}{R35}.
\bibitem[{Engl et~al.(1996)Engl, Hanke and Neubauer}]{engl1996regularization}
\bibinfo{author}{Engl, H.W.}, \bibinfo{author}{Hanke, M.},
  \bibinfo{author}{Neubauer, A.}, \bibinfo{year}{1996}.
\newblock \bibinfo{title}{Regularization of inverse problems}. volume
  \bibinfo{volume}{375}.
\newblock \bibinfo{publisher}{Springer Science \& Business Media}.
\bibitem[{Finocchiaro et~al.(2021)Finocchiaro, Sheikh, Leone, Westaby,
  Mazzarotto, Pantazis, Ferrantini, Sacconi, Papadakis, Sharma
  et~al.}]{finocchiaro2021arrhythmogenic}
\bibinfo{author}{Finocchiaro, G.}, \bibinfo{author}{Sheikh, N.},
  \bibinfo{author}{Leone, O.}, \bibinfo{author}{Westaby, J.},
  \bibinfo{author}{Mazzarotto, F.}, \bibinfo{author}{Pantazis, A.},
  \bibinfo{author}{Ferrantini, C.}, \bibinfo{author}{Sacconi, L.},
  \bibinfo{author}{Papadakis, M.}, \bibinfo{author}{Sharma, S.}, et~al.,
  \bibinfo{year}{2021}.
\newblock \bibinfo{title}{Arrhythmogenic potential of myocardial disarray in
  hypertrophic cardiomyopathy: genetic basis, functional consequences and
  relation to sudden cardiac death}.
\newblock \bibinfo{journal}{EP Europace} \bibinfo{volume}{23},
  \bibinfo{pages}{985--995}.
\bibitem[{Finsberg et~al.(2018)Finsberg, Balaban, Ross, H{\aa}land, Odland,
  Sundnes and Wall}]{finsberg}
\bibinfo{author}{Finsberg, H.}, \bibinfo{author}{Balaban, G.},
  \bibinfo{author}{Ross, S.}, \bibinfo{author}{H{\aa}land, T.F.},
  \bibinfo{author}{Odland, H.}, \bibinfo{author}{Sundnes, J.},
  \bibinfo{author}{Wall, S.}, \bibinfo{year}{2018}.
\newblock \bibinfo{title}{Estimating cardiac contraction through high
  resolution data assimilation of a personalized mechanical model}.
\newblock \bibinfo{journal}{Journal of computational science}
  \bibinfo{volume}{24}, \bibinfo{pages}{85--90}.
\bibitem[{Gander et~al.(2021)Gander, Krause, Multerer and
  Pezzuto}]{Gander2021UQ}
\bibinfo{author}{Gander, L.}, \bibinfo{author}{Krause, R.},
  \bibinfo{author}{Multerer, M.}, \bibinfo{author}{Pezzuto, S.},
  \bibinfo{year}{2021}.
\newblock \bibinfo{title}{Space-time shape uncertainty in the forward and
  inverse problem of electrocardiography}.
\newblock \bibinfo{journal}{International Journal for Numerical Methods in
  Biomedical Engineering} \bibinfo{volume}{37}.
\newblock \DOIprefix\doi{10.1002/cnm.3522},
  \href{http://arxiv.org/abs/2010.16104}{{\tt arXiv:2010.16104}}.
\bibitem[{Gjerald et~al.(2015)Gjerald, Hake, Pezzuto, Sundnes and
  Wall}]{Sjur2015Stacom}
\bibinfo{author}{Gjerald, S.}, \bibinfo{author}{Hake, J.},
  \bibinfo{author}{Pezzuto, S.}, \bibinfo{author}{Sundnes, J.},
  \bibinfo{author}{Wall, S.T.}, \bibinfo{year}{2015}.
\newblock \bibinfo{title}{Patient-specific parameter estimation for a
  transversely isotropic active strain model of left ventricular mechanics},
  in: \bibinfo{editor}{Camara, O.}, \bibinfo{editor}{Mansi, T.},
  \bibinfo{editor}{Pop, M.}, \bibinfo{editor}{Rhode, K.},
  \bibinfo{editor}{Sermesant, M.}, \bibinfo{editor}{Young, A.} (Eds.),
  \bibinfo{booktitle}{Statistical Atlases and Computational Models of the
  Heart: Imaging and Modelling Challenges}, \bibinfo{publisher}{Springer},
  \bibinfo{address}{Cham}. pp. \bibinfo{pages}{93--104}.
\newblock \DOIprefix\doi{10.1007/978-3-319-14678-2_10}.
\bibitem[{Guccione et~al.(1991)Guccione, McCulloch and
  Waldman}]{guccione1991passive}
\bibinfo{author}{Guccione, J.M.}, \bibinfo{author}{McCulloch, A.D.},
  \bibinfo{author}{Waldman, L.K.}, \bibinfo{year}{1991}.
\newblock \bibinfo{title}{{Passive Material Properties of Intact Ventricular
  Myocardium Determined From a Cylindrical Model}}.
\newblock \bibinfo{journal}{Journal of Biomechanical Engineering}
  \bibinfo{volume}{113}, \bibinfo{pages}{42--55}.
\newblock \URLprefix \url{https://doi.org/10.1115/1.2894084},
  \DOIprefix\doi{10.1115/1.2894084}.
\bibitem[{Ham et~al.(2023)Ham, Kelly, Mitchell, Cotter, Kirby, Sagiyama,
  Bouziani, Vorderwuelbecke, Gregory, Betteridge, Shapero, Nixon-Hill, Ward,
  Farrell, Brubeck, Marsden, Gibson, Homolya, Sun, McRae, Luporini, Gregory,
  Lange, Funke, Rathgeber, Bercea and Markall}]{FiredrakeUserManual}
\bibinfo{author}{Ham, D.A.}, \bibinfo{author}{Kelly, P.H.J.},
  \bibinfo{author}{Mitchell, L.}, \bibinfo{author}{Cotter, C.J.},
  \bibinfo{author}{Kirby, R.C.}, \bibinfo{author}{Sagiyama, K.},
  \bibinfo{author}{Bouziani, N.}, \bibinfo{author}{Vorderwuelbecke, S.},
  \bibinfo{author}{Gregory, T.J.}, \bibinfo{author}{Betteridge, J.},
  \bibinfo{author}{Shapero, D.R.}, \bibinfo{author}{Nixon-Hill, R.W.},
  \bibinfo{author}{Ward, C.J.}, \bibinfo{author}{Farrell, P.E.},
  \bibinfo{author}{Brubeck, P.D.}, \bibinfo{author}{Marsden, I.},
  \bibinfo{author}{Gibson, T.H.}, \bibinfo{author}{Homolya, M.},
  \bibinfo{author}{Sun, T.}, \bibinfo{author}{McRae, A.T.T.},
  \bibinfo{author}{Luporini, F.}, \bibinfo{author}{Gregory, A.},
  \bibinfo{author}{Lange, M.}, \bibinfo{author}{Funke, S.W.},
  \bibinfo{author}{Rathgeber, F.}, \bibinfo{author}{Bercea, G.T.},
  \bibinfo{author}{Markall, G.R.}, \bibinfo{year}{2023}.
\newblock \bibinfo{title}{Firedrake User Manual}. \bibinfo{edition}{first
  edition} ed.
\newblock \bibinfo{organization}{Imperial College London and University of
  Oxford and Baylor University and University of Washington}.
\newblock \DOIprefix\doi{10.25561/104839}.
\bibitem[{Holzapfel and Ogden(2009)}]{holzapfel2009constitutive}
\bibinfo{author}{Holzapfel, G.A.}, \bibinfo{author}{Ogden, R.W.},
  \bibinfo{year}{2009}.
\newblock \bibinfo{title}{Constitutive modelling of passive myocardium: a
  structurally based framework for material characterization}.
\newblock \bibinfo{journal}{Philosophical Transactions of the Royal Society A:
  Mathematical, Physical and Engineering Sciences} \bibinfo{volume}{367},
  \bibinfo{pages}{3445--3475}.
\bibitem[{Hughes and Marsden(1983)}]{hughes1983mathematical}
\bibinfo{author}{Hughes, T.}, \bibinfo{author}{Marsden, J.},
  \bibinfo{year}{1983}.
\newblock \bibinfo{title}{Mathematical foundations of elasticity}.
\newblock \bibinfo{publisher}{Citeseer}.
\bibitem[{Kovacheva et~al.(2021)Kovacheva, Th{\"a}mer, Fritz, Seemann, Ochs,
  D{\"o}ssel and Loewe}]{kovacheva}
\bibinfo{author}{Kovacheva, E.}, \bibinfo{author}{Th{\"a}mer, L.},
  \bibinfo{author}{Fritz, T.}, \bibinfo{author}{Seemann, G.},
  \bibinfo{author}{Ochs, M.}, \bibinfo{author}{D{\"o}ssel, O.},
  \bibinfo{author}{Loewe, A.}, \bibinfo{year}{2021}.
\newblock \bibinfo{title}{Estimating cardiac active tension from wall
  motion—an inverse problem of cardiac biomechanics}.
\newblock \bibinfo{journal}{International Journal for Numerical Methods in
  Biomedical Engineering} \bibinfo{volume}{37}, \bibinfo{pages}{e3448}.
\bibitem[{Lee et~al.(2007)Lee, Ingrassia, Costa, Holmes, Konofagou
  et~al.}]{lee2007theoretical}
\bibinfo{author}{Lee, W.}, \bibinfo{author}{Ingrassia, C.},
  \bibinfo{author}{Costa, K.}, \bibinfo{author}{Holmes, J.},
  \bibinfo{author}{Konofagou, E.}, et~al., \bibinfo{year}{2007}.
\newblock \bibinfo{title}{Theoretical quality assessment of myocardial
  elastography with in vivo validation}.
\newblock \bibinfo{journal}{ieee transactions on ultrasonics, ferroelectrics,
  and frequency control} \bibinfo{volume}{54}, \bibinfo{pages}{2233--2245}.
\bibitem[{Lions(1971)}]{lions1971}
\bibinfo{author}{Lions, J.L.}, \bibinfo{year}{1971}.
\newblock \bibinfo{title}{Optimal control of systems governed by partial
  differential equations}. volume \bibinfo{volume}{170}.
\newblock \bibinfo{publisher}{Springer}.
\bibitem[{Lubrecht et~al.(2021)Lubrecht, Grandits, Gharaviri, Schotten, Pock,
  Plank, Krause, Auricchio and Pezzuto}]{Lubrecht2021PIEMAP}
\bibinfo{author}{Lubrecht, J.M.}, \bibinfo{author}{Grandits, T.},
  \bibinfo{author}{Gharaviri, A.}, \bibinfo{author}{Schotten, U.},
  \bibinfo{author}{Pock, T.}, \bibinfo{author}{Plank, G.},
  \bibinfo{author}{Krause, R.}, \bibinfo{author}{Auricchio, A.},
  \bibinfo{author}{Pezzuto, S.}, \bibinfo{year}{2021}.
\newblock \bibinfo{title}{Automatic reconstruction of the left atrium
  activation from sparse intracardiac contact recordings by inverse estimate of
  fiber structure and anisotropic conduction in a patient-specific model}.
\newblock \bibinfo{journal}{EP Europace} \bibinfo{volume}{23},
  \bibinfo{pages}{i63--i70}.
\newblock \DOIprefix\doi{10.1093/europace/euaa392}.
\bibitem[{Maffessanti et~al.(2020)Maffessanti, Jadczyk, Kurzelowski, Regoli,
  Caputo, Conte, Go{\l}ba, Biernat, Wilczek, D{\k{a}}browska, Pezzuto,
  Moccetti, Krause, Wojakowski, Prinzen and Auricchio}]{Maffessanti2020scar}
\bibinfo{author}{Maffessanti, F.}, \bibinfo{author}{Jadczyk, T.},
  \bibinfo{author}{Kurzelowski, R.}, \bibinfo{author}{Regoli, F.},
  \bibinfo{author}{Caputo, M.L.}, \bibinfo{author}{Conte, G.},
  \bibinfo{author}{Go{\l}ba, K.S.}, \bibinfo{author}{Biernat, J.},
  \bibinfo{author}{Wilczek, J.}, \bibinfo{author}{D{\k{a}}browska, M.},
  \bibinfo{author}{Pezzuto, S.}, \bibinfo{author}{Moccetti, T.},
  \bibinfo{author}{Krause, R.}, \bibinfo{author}{Wojakowski, W.},
  \bibinfo{author}{Prinzen, F.W.}, \bibinfo{author}{Auricchio, A.},
  \bibinfo{year}{2020}.
\newblock \bibinfo{title}{The influence of scar on the spatio-temporal
  relationship between electrical and mechanical activation in heart failure
  patients}.
\newblock \bibinfo{journal}{EP Europace} \bibinfo{volume}{22},
  \bibinfo{pages}{777--786}.
\newblock \DOIprefix\doi{10.1093/europace/euz346}.
\bibitem[{Manzoni et~al.()Manzoni, Quarteroni and Salsa}]{manzoni2021optimal}
\bibinfo{author}{Manzoni, A.}, \bibinfo{author}{Quarteroni, A.},
  \bibinfo{author}{Salsa, S.}, .
\newblock \bibinfo{title}{Optimal control of partial differential equations}.
\newblock \bibinfo{publisher}{Springer}.
\bibitem[{Mitusch et~al.(2019)Mitusch, Funke and Dokken}]{mitusch2019dolfin}
\bibinfo{author}{Mitusch, S.}, \bibinfo{author}{Funke, S.},
  \bibinfo{author}{Dokken, J.}, \bibinfo{year}{2019}.
\newblock \bibinfo{title}{dolfin-adjoint 2018.1: automated adjoints for fenics
  and firedrake}.
\newblock \bibinfo{journal}{Journal of Open Source Software}
  \bibinfo{volume}{4}, \bibinfo{pages}{1292}.
\bibitem[{Mojsejenko et~al.(2015)Mojsejenko, McGarvey, Dorsey, Gorman, Burdick,
  Pilla, Gorman and Wenk}]{mojsejenko2015estimating}
\bibinfo{author}{Mojsejenko, D.}, \bibinfo{author}{McGarvey, J.},
  \bibinfo{author}{Dorsey, S.}, \bibinfo{author}{Gorman, J.},
  \bibinfo{author}{Burdick, J.}, \bibinfo{author}{Pilla, J.},
  \bibinfo{author}{Gorman, R.}, \bibinfo{author}{Wenk, J.},
  \bibinfo{year}{2015}.
\newblock \bibinfo{title}{Estimating passive mechanical properties in a
  myocardial infarction using mri and finite element simulations}.
\newblock \bibinfo{journal}{Biomechanics and modeling in mechanobiology}
  \bibinfo{volume}{14}, \bibinfo{pages}{633--647}.
\bibitem[{Moulton et~al.(1995)Moulton, Creswell, Actis, Myers, Vannier, Szabo
  and Pasque}]{moulton1995inverse}
\bibinfo{author}{Moulton, M.}, \bibinfo{author}{Creswell, L.},
  \bibinfo{author}{Actis, R.}, \bibinfo{author}{Myers, K.},
  \bibinfo{author}{Vannier, M.}, \bibinfo{author}{Szabo, B.},
  \bibinfo{author}{Pasque, M.}, \bibinfo{year}{1995}.
\newblock \bibinfo{title}{An inverse approach to determining myocardial
  material properties}.
\newblock \bibinfo{journal}{Journal of biomechanics} \bibinfo{volume}{28},
  \bibinfo{pages}{935--948}.
\bibitem[{Nair et~al.(2007)Nair, Taggart and Vetter}]{nair2007optimizing}
\bibinfo{author}{Nair, A.}, \bibinfo{author}{Taggart, D.},
  \bibinfo{author}{Vetter, F.}, \bibinfo{year}{2007}.
\newblock \bibinfo{title}{Optimizing cardiac material parameters with a genetic
  algorithm}.
\newblock \bibinfo{journal}{Journal of biomechanics} \bibinfo{volume}{40},
  \bibinfo{pages}{1646--1650}.
\bibitem[{Nocedal and Wright(1999)}]{nocedal1999numerical}
\bibinfo{author}{Nocedal, J.}, \bibinfo{author}{Wright, S.J.},
  \bibinfo{year}{1999}.
\newblock \bibinfo{title}{Numerical optimization}.
\newblock \bibinfo{publisher}{Springer}.
\bibitem[{Quaglino et~al.(2019)Quaglino, Pezzuto and Krause}]{Quaglino2019MFMC}
\bibinfo{author}{Quaglino, A.}, \bibinfo{author}{Pezzuto, S.},
  \bibinfo{author}{Krause, R.}, \bibinfo{year}{2019}.
\newblock \bibinfo{title}{High-dimensional and higher-order multifidelity monte
  carlo estimators}.
\newblock \bibinfo{journal}{Journal of Computational Physics}
  \bibinfo{volume}{388}, \bibinfo{pages}{300--315}.
\newblock \DOIprefix\doi{10.1016/j.jcp.2019.03.026}.
\bibitem[{Regazzoni et~al.(2022)Regazzoni, Salvador, Dede and
  Quarteroni}]{regazzoni2022machine}
\bibinfo{author}{Regazzoni, F.}, \bibinfo{author}{Salvador, M.},
  \bibinfo{author}{Dede, L.}, \bibinfo{author}{Quarteroni, A.},
  \bibinfo{year}{2022}.
\newblock \bibinfo{title}{A machine learning method for real-time numerical
  simulations of cardiac electromechanics}.
\newblock \bibinfo{journal}{Computer methods in applied mechanics and
  engineering} \bibinfo{volume}{393}, \bibinfo{pages}{114825}.
\bibitem[{Riccobelli and Ambrosi(2019)}]{riccobelli2019activation}
\bibinfo{author}{Riccobelli, D.}, \bibinfo{author}{Ambrosi, D.},
  \bibinfo{year}{2019}.
\newblock \bibinfo{title}{Activation of a muscle as a mapping of stress--strain
  curves}.
\newblock \bibinfo{journal}{Extreme Mechanics Letters} \bibinfo{volume}{28},
  \bibinfo{pages}{37--42}.
\bibitem[{Rudin et~al.(1992)Rudin, Osher and Fatemi}]{rudin1992nonlinear}
\bibinfo{author}{Rudin, L.I.}, \bibinfo{author}{Osher, S.},
  \bibinfo{author}{Fatemi, E.}, \bibinfo{year}{1992}.
\newblock \bibinfo{title}{Nonlinear total variation based noise removal
  algorithms}.
\newblock \bibinfo{journal}{Physica D: nonlinear phenomena}
  \bibinfo{volume}{60}, \bibinfo{pages}{259--268}.
\bibitem[{Ruiz-Herrera et~al.(2022)Ruiz-Herrera, Grandits, Plank, Perdikaris,
  Sahli~Costabal and Pezzuto}]{RuizPINN2022}
\bibinfo{author}{Ruiz-Herrera, C.}, \bibinfo{author}{Grandits, T.},
  \bibinfo{author}{Plank, G.}, \bibinfo{author}{Perdikaris, P.},
  \bibinfo{author}{Sahli~Costabal, F.}, \bibinfo{author}{Pezzuto, S.},
  \bibinfo{year}{2022}.
\newblock \bibinfo{title}{Physics-informed neural networks to learn cardiac
  fiber orientation from multiple electro\-anatomical maps}.
\newblock \bibinfo{journal}{Engineering with Computers}
  \DOIprefix\doi{10.1007/s00366-022-01709-3},
  \href{http://arxiv.org/abs/2201.12362}{{\tt arXiv:2201.12362}}.
\bibitem[{Sack et~al.(2016)Sack, Davies, Guccione and
  Franz}]{sack2016personalised}
\bibinfo{author}{Sack, K.}, \bibinfo{author}{Davies, N.},
  \bibinfo{author}{Guccione, J.}, \bibinfo{author}{Franz, T.},
  \bibinfo{year}{2016}.
\newblock \bibinfo{title}{Personalised computational cardiology:
  patient-specific modelling in cardiac mechanics and biomaterial injection
  therapies for myocardial infarction}.
\newblock \bibinfo{journal}{Heart failure reviews} \bibinfo{volume}{21},
  \bibinfo{pages}{815--826}.
\bibitem[{Schr{\"o}der and Neff(2003)}]{schroder2003invariant}
\bibinfo{author}{Schr{\"o}der, J.}, \bibinfo{author}{Neff, P.},
  \bibinfo{year}{2003}.
\newblock \bibinfo{title}{Invariant formulation of hyperelastic transverse
  isotropy based on polyconvex free energy functions}.
\newblock \bibinfo{journal}{International journal of solids and structures}
  \bibinfo{volume}{40}, \bibinfo{pages}{401--445}.
\bibitem[{Sermesant et~al.(2006)Sermesant, Moireau, Camara, Sainte-Marie,
  Andriantsimiavona, Cimrman, Hill, Chapelle and Razavi}]{sermesant2006cardiac}
\bibinfo{author}{Sermesant, M.}, \bibinfo{author}{Moireau, P.},
  \bibinfo{author}{Camara, O.}, \bibinfo{author}{Sainte-Marie, J.},
  \bibinfo{author}{Andriantsimiavona, R.}, \bibinfo{author}{Cimrman, R.},
  \bibinfo{author}{Hill, D.}, \bibinfo{author}{Chapelle, D.},
  \bibinfo{author}{Razavi, R.}, \bibinfo{year}{2006}.
\newblock \bibinfo{title}{Cardiac function estimation from mri using a heart
  model and data assimilation: advances and difficulties}.
\newblock \bibinfo{journal}{Medical image analysis} \bibinfo{volume}{10},
  \bibinfo{pages}{642--656}.
\bibitem[{Sun et~al.(2009)Sun, Stander, Jhun, Zhang, Suzuki, Wang, Saeed,
  Wallace, Tseng, Baker et~al.}]{sun2009computationally}
\bibinfo{author}{Sun, K.}, \bibinfo{author}{Stander, N.},
  \bibinfo{author}{Jhun, C.}, \bibinfo{author}{Zhang, Z.},
  \bibinfo{author}{Suzuki, T.}, \bibinfo{author}{Wang, G.Y.},
  \bibinfo{author}{Saeed, M.}, \bibinfo{author}{Wallace, A.},
  \bibinfo{author}{Tseng, E.}, \bibinfo{author}{Baker, A.}, et~al.,
  \bibinfo{year}{2009}.
\newblock \bibinfo{title}{A computationally efficient formal optimization of
  regional myocardial contractility in a sheep with left ventricular aneurysm}.
\newblock \bibinfo{journal}{J.~Biomech.~Eng} \bibinfo{volume}{131},
  \bibinfo{pages}{111001}.
\bibitem[{Tortora and Derrickson(2014)}]{tortora2014anatomy}
\bibinfo{author}{Tortora, G.}, \bibinfo{author}{Derrickson, B.},
  \bibinfo{year}{2014}.
\newblock \bibinfo{title}{Anatomy \& physiology}.
\newblock \bibinfo{publisher}{Wiley India Pvt Limited}.
\bibitem[{Tr{\"o}ltzsch(2010)}]{troltzsch2010optimal}
\bibinfo{author}{Tr{\"o}ltzsch, F.}, \bibinfo{year}{2010}.
\newblock \bibinfo{title}{Optimal control of partial differential equations}.
\newblock \bibinfo{journal}{Graduate studies in mathematics}
  \bibinfo{volume}{112}, \bibinfo{pages}{399}.
\bibitem[{Wang et~al.(2009)Wang, Lam, Ennis, Cowan, Young and
  Nash}]{wang2009modelling}
\bibinfo{author}{Wang, V.}, \bibinfo{author}{Lam, H.}, \bibinfo{author}{Ennis,
  D.}, \bibinfo{author}{Cowan, B.}, \bibinfo{author}{Young, A.},
  \bibinfo{author}{Nash, M.}, \bibinfo{year}{2009}.
\newblock \bibinfo{title}{Modelling passive diastolic mechanics with
  quantitative mri of cardiac structure and function}.
\newblock \bibinfo{journal}{Medical image analysis} \bibinfo{volume}{13},
  \bibinfo{pages}{773--784}.
\bibitem[{Zervantonakis et~al.(2007)Zervantonakis, Fung-Kee-Fung, Lee and
  Konofagou}]{zervantonakis2007novel}
\bibinfo{author}{Zervantonakis, I.}, \bibinfo{author}{Fung-Kee-Fung, S.},
  \bibinfo{author}{Lee, W.}, \bibinfo{author}{Konofagou, E.},
  \bibinfo{year}{2007}.
\newblock \bibinfo{title}{A novel, view-independent method for strain mapping
  in myocardial elastography: eliminating angle and centroid dependence}.
\newblock \bibinfo{journal}{Physics in Medicine \& Biology}
  \bibinfo{volume}{52}, \bibinfo{pages}{4063}.
\bibitem[{Zhu et~al.(1997)Zhu, Byrd, Lu and Nocedal}]{zhu1997algorithm}
\bibinfo{author}{Zhu, C.}, \bibinfo{author}{Byrd, R.H.}, \bibinfo{author}{Lu,
  P.}, \bibinfo{author}{Nocedal, J.}, \bibinfo{year}{1997}.
\newblock \bibinfo{title}{Algorithm 778: L-bfgs-b: Fortran subroutines for
  large-scale bound-constrained optimization}.
\newblock \bibinfo{journal}{ACM Transactions on mathematical software (TOMS)}
  \bibinfo{volume}{23}, \bibinfo{pages}{550--560}.

\end{thebibliography}

\appendix

\section{Boundary conditions and system observability}
\label{appendix::A}
\begin{figure}[h]
	\centering
	\includegraphics[width=0.135\textwidth]{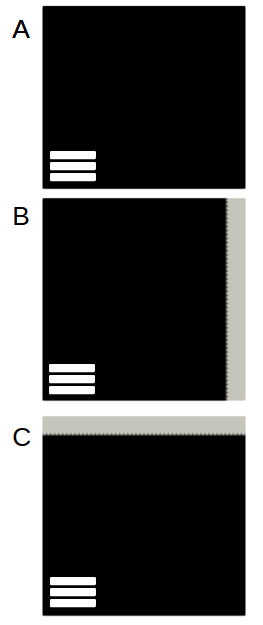}
    \includegraphics[width=0.32\textwidth]{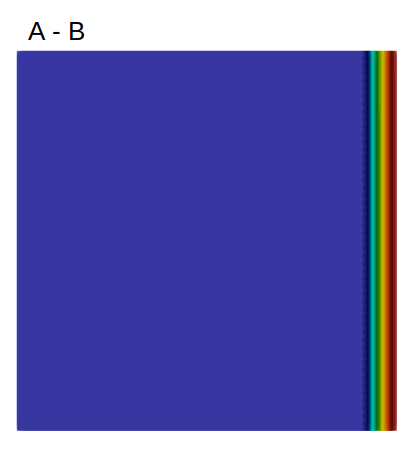}
	\includegraphics[width=0.32\textwidth]{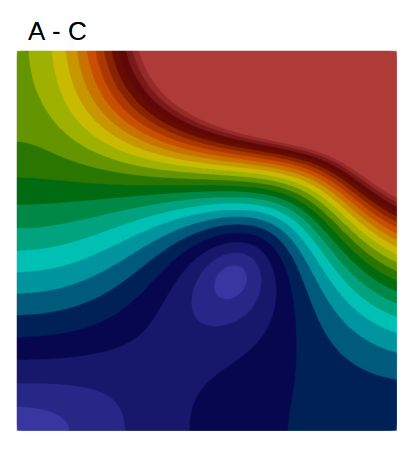}
	\includegraphics[width=0.16\textwidth]{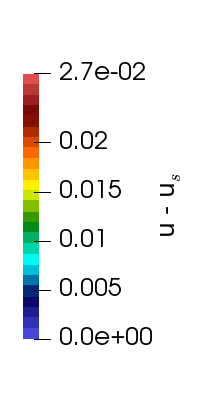}
	\caption{Effect of a scar located at the free-boundary on the displacement field. (Left) Sketch of the three different scenarios (A)-(C) reproduced in the tests, \rev{with healthy tissue in black and scar in gray. Fibers direction is indicated in the lower left corner. 
    (Middle) Difference in magnitude between the displacement field of the case (A) and the case (B). The discrepancy is only limited to the scarred region.
    (Right) Difference in magnitude between the displacement field of the case (A) and the case (C). The discrepancy is larger and more distributed.}}
	\label{fig::observability}
\end{figure}

As shown in Section \ref{subsec::SA}, the use of the regularizing term $\mRL$ rises the issue on how the system observability may be affected by the choice of the boundary conditions.
Here, we provide arguments in support of the hypothesis that the poor reconstruction of contractility, that we obtain with $\mRL$ and a stress-free condition on the boundary perpendicular to the fibers direction, is imputable to a not fully observable system.
To this end, we show how the displacement field can be modified by the presence of a \rev{rectangular} scar located at the stress-free boundary, both for fibers arrange\rev{d} perpendicularly and parallel to that boundary, see Fig.~\ref{fig::observability} \rev{(Left) (B) and (C) respectively}. We notice that when the scar is located in a region close to the boundary parallel to the fiber orientation $(C)$, the displacement field is more affected, both in terms of magnitude and spatial localization, than the one obtained when the scar is located in a region perpendicular $(B)$ to the fiber orientation.
Accordingly, $\Delta^i = \Vert \bu - \bu_s^i \Vert/\Vert \bu \Vert$ is significantly larger in the latter case (i.e. $\Delta^{\parallel}/\Delta^{\perp} \simeq 10$). \rev{Here with \rev{$\bu_s^i$} we denote the displacement field for the configuration presenting the scar either perpendicular ($i = \perp$) or parallel ($i = \parallel$) to the fiber orientation and with $\bu$ the displacement field of the configuration without the scar.} 
Therefore, a scar positioned on the stress-free boundary orthogonal to the fibers does not introduce appreciable modification in the displacement field with respect to the solution obtained in case of healthy tissue $(A)$. In other terms, Fig. \ref{fig::observability} (center) demonstrates that different contractility functions can result in nearly identical displacement fields. We notice that this pathological condition does not emerge when the scar is located at a stress-free border parallel to the fibers direction. In the former case, the choice of the regularizing term plays a fundamental role in restricting the field of admissible $\alpha(x,y)$ for a given observation $\bu_o$, by penalizing non-physical features of the control variable. As so, the choice of $\mRL$ results inappropriate for the above described problem setting and generate\rev{s} inaccurate reconstructions of the contractility as shown in Fig.~\ref{fig::fib_orient}.

\end{document}